\documentclass[%
pra,
twocolumn,
showpacs,
letterpaper,
reprint,
longbibliography,
amsmath,
amssymb,
floatfix,
superscriptaddress
]{revtex4-1}

\pdfoutput=1

 \usepackage[ruled,vlined]{algorithm2e}
 \usepackage{natbib}
 \usepackage{dcolumn}
 
 \usepackage{graphicx,color}
 \usepackage{array}

 \PassOptionsToPackage{caption=false}{subfig}
 \usepackage{subfig}
 \usepackage{url}

\newcommand{\ic}{\ensuremath{\text{I}_\text{c}}}
\newcommand{\ip}{\ensuremath{\text{I}_\text{p}}}
\newcommand{\lqu}{\ensuremath{\text{L}_\text{q}}}
\newcommand{\lcjj}{\ensuremath{\text{L}_\text{cjj}}}

\begin{document}

\title{A scalable control system for a superconducting adiabatic quantum optimization processor}

\author{M.~W.~Johnson}
\affiliation{D-Wave Systems Inc., 100-4401
  Still Creek Dr., Burnaby, BC V5C 6G9 Canada }  
\email{mwjohnson@dwavesys.com}


\author{P.~Bunyk}
\affiliation{D-Wave Systems Inc., 100-4401
  Still Creek Dr., Burnaby, BC V5C 6G9 Canada }
\author{F.~Maibaum}
\affiliation{Physikalisch
  Technische Bundesanstalt, Bundesallee 100, 38116 Braunschweig,
  Germany}
\author{E.~Tolkacheva}
\author{A.~J.~Berkley}
\author{E.~M.~Chapple}
\author{R.~Harris}
\author{J.~Johansson}
\author{T.~Lanting}
\author{I.~Perminov}
\author{E.~Ladizinsky}
\author{T.~Oh}
\author{G.~Rose}
\affiliation{D-Wave Systems Inc., 100-4401
  Still Creek Dr., Burnaby, BC V5C 6G9 Canada }

\markboth{A scalable control system for a superconducting AQO processor}{}

\begin{abstract}
We have designed, fabricated and operated a scalable system for
applying independently programmable time-independent, and limited
time-dependent flux biases to control superconducting devices in an
integrated circuit. Here we report on the operation of a system
designed to supply 64 flux biases to devices in a circuit designed to
be a unit cell for a superconducting adiabatic quantum optimization (AQO)
system. The system requires six digital address lines, two power lines,
and a handful of global analog lines.
\end{abstract}

\maketitle

\section{Introduction}
Several proposals for how one might implement a
quantum computer now exist. One of these is based on enabling
adiabatic quantum optimization algorithms in networks of
superconducting flux qubits connected via tunable coupling devices
\citep{kaminsky}. Flux qubits can be manipulated by applying magnetic
flux via currents along inductively coupled control lines. This can be
accomplished with one analog control line per device driven by room
temperature current sources and routed, through appropriate filtering,
down to the target device on chip.

Beyond the scale of a few dozens of such qubits the
\emph{one-analog-line-per-device} approach becomes impractical.
Hundreds of qubits could require thousands of wires, each subject to
filtering, cross-talk, and thermal requirements so as to minimize
disturbance of the thermal and electromagnetic environment of the
targeted qubits, which are operated at milliKelvin temperatures. We
require an approach that does not use so many wires.

One advantage of using superconductor based qubits is the existence of
a compatible classical digital and mixed signal electronics technology
based on the manipulation of single flux quanta (SFQ)
\cite{likharev,hurrell}.  The ability to manufacture classical control
circuitry \cite{abelson,silver,hypres} on the same chip, with the same
fabrication technology as is used in construction of the qubits,
addresses many of the thermal and electromagnetic compatibility
requirements faced in integrating control circuitry with such a
processor.  The idea of using SFQ circuitry to control flux qubits is
not new, and has investigated by a number of researchers
\cite{castellano,Xingxiang,deCastro,orlando,semenov,berggren,seva}.

We present here a description of a functioning system of on-chip
Programmable Magnetic Memory (PMM) designed to manipulate the
parameters and state of superconducting flux qubits and tunable
couplers, in such a way as to overcome the scalability limitations of
the \emph{one-analog-line-per-device} paradigm.  This system
comprises three key parts.

\begin{figure}
\includegraphics[width=3.3in]{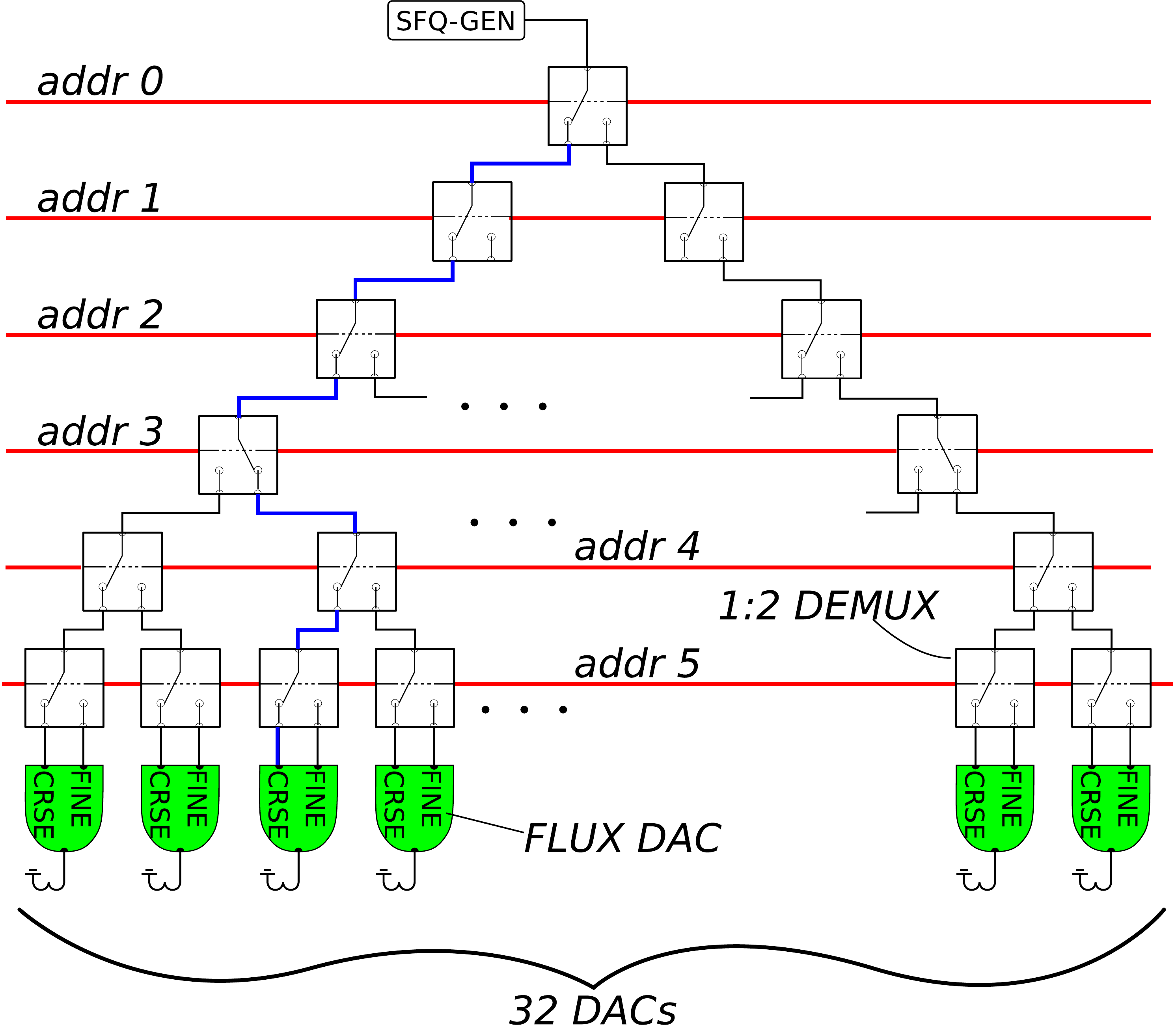}
\caption{A 1:32 demultiplexer tree terminating in two-stage multiple
  flux quantum DACs.  The last address selects between the COARSE and
  FINE stages within a DAC.  Two such trees were implemented for the
  64 DAC circuit reported here.}
\label{fig:tree}
\end{figure}

The first of these is a SFQ demultiplexer used as an addressing
system.  It is constructed as a binary tree of $2^N-1$ 1:2 SFQ
demultiplexer gates as shown in Fig.~\ref{fig:tree}.  For the specific
design discussed here, the number of address lines $N$ is 6.  This
demultiplexer allows many devices to be addressed using only a few
address lines.

The second part is a set of digital-to-analog converters (DACs),
located at the leaves of the address tree.  These DACs comprise
storage inductors that can hold an integer number of single magnetic
flux quanta $(\Phi_0 = h/2e)$.  Their digital input are single flux
quanta, and their analog output are the stored flux, which can be
coupled into a target device.  The magnitude of this output flux is
proportional to the number of stored flux quanta. Each DAC has two
such storage inductors, a COARSE stage, and a FINE stage, named for
the relative strength with which their output flux is coupled to the
target device.  In our architecture, the output of these DACs is
static.

The third part is a method for converting the static output of a DAC
to a time-dependent signal. This is achieved by coupling the output of
the DAC into a variable gain element, equivalent to the tunable
coupler described elsewhere \cite{coupler-paper}.  An analog line carrying a
time-dependent current is coupled to the target device via the
variable gain element. This approach is useful in the types of
circuits of interest here because single analog lines can be
shared among large numbers of devices that need the same functional
dependence on time, but may require individual tunability of the gain
and offset.

We designed, fabricated, and operated integrated circuits comprising
this type of control system architecture. One such circuit includes
eight superconducting rf-SQUID flux qubits as described in
\cite{robust-scalable}.  Though not discussed here, the state of each
qubit was read out via a Quantum Flux Parametron (QFP) latch or buffer
which was in turn read by an x-y addressible dc-SQUID array as
discussed in detail in \cite{readout}.  Each rf-SQUID flux qubit has
inductive ports coupled to five different DACs, and 24 compound
Josephson junction (CJJ) rf-SQUID couplers \cite{coupler-paper}, each
of which is coupled to a single DAC.  Thus, this circuit required $8
\times 5 + 24 = 64$ DACs. The particular control circuit described
here comprises two 1:32 demultiplexers with six shared address lines
and two separate power lines. This circuit included 1,538 junctions
ranging in size from a minimum of 0.6 $\mu$m diameter (32 of them) to
a maximum of 4~$\mu$m in diameter.

The paper is organized as follows: Requirements on control circuitry
derived from the devices, architecture and operating procedures in
superconducting adiabatic quantum optimization systems are discussed
in section~\ref{sect:require}. The specific control circuitry
architecture is discussed in section~\ref{sect:arch}. Data
demonstrating the performance relative to requirements is presented in
section~\ref{sect:data}. Conclusions are presented in
section~\ref{sect:conc}.

\begin{figure}[t]
\includegraphics[width=3.3in]{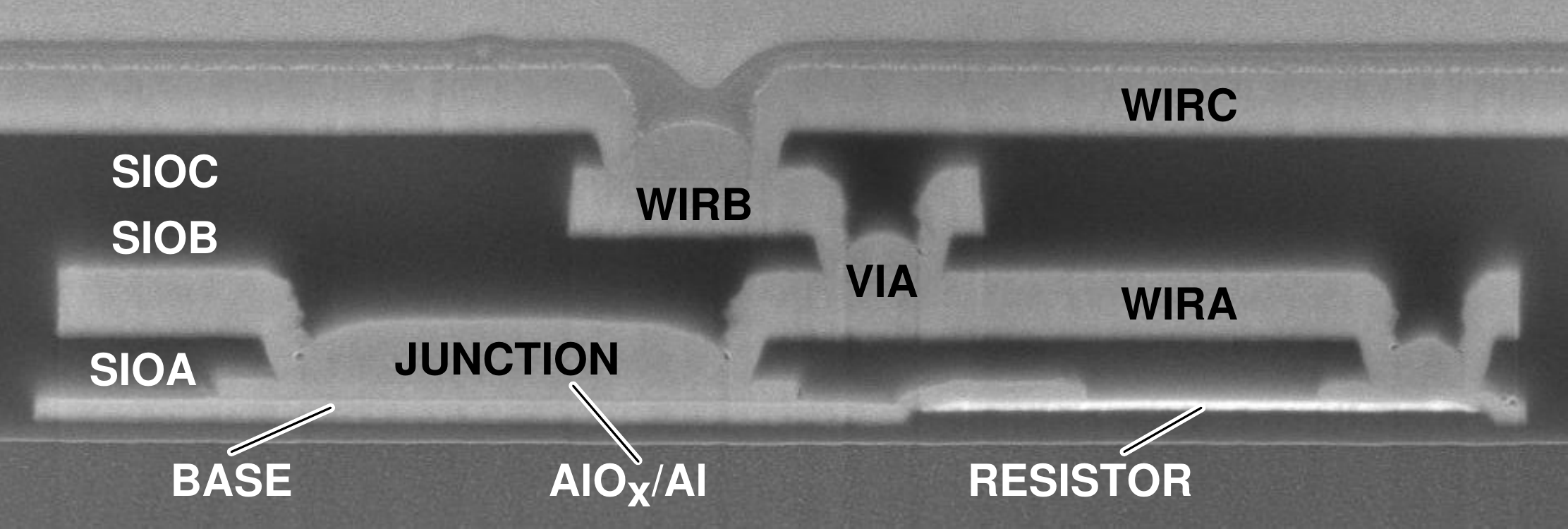}
\caption{FIB cut SEM cross-section for the process used to fabricate
  the circuits described here.}
\label{fig:xsect}
\end{figure}

The measurements reported in section~\ref{sect:data} were performed on
chips fabricated in a four Nb layer superconducting process employing
a standard Nb/AlOx/Nb trilayer, a TiPt resistor layer, and planarized
PECVD SiO$_2$ dielectric layers. Design rules included 0.25 $\mu m$
lines and spaces for wiring layers and a minimum junction diameter of
0.6 $\mu m$.  A sample process cross section is shown in
Fig.~\ref{fig:xsect}.  

\section{Magnetic Memory Requirements}
\label{sect:require}

Our intent is to embody a specific quantum algorithm in hardware. This
algorithm, known as adiabatic quantum optimization (AQO), is a novel
approach for solving combinatorial optimization
problems\cite{farhi2001,steffen}. Unlike the incumbent techniques for
such problems, such as simulated annealing or genetic algorithms, AQO
algorithms include procedures that are explicitly quantum
mechanical. The requirement to provide the quantum mechanical
resources necessary for running this algorithm places unusual
constraints on processor systems and components.

\begin{algorithm}
  \SetAlgoLined
  \SetKwInOut{Input}{Input}\SetKwInOut{Output}{Output}
  \SetKw{load}{load}
  \SetKw{set}{set}
  \SetKw{wait}{wait}
  \Input{A run-time $t_f$; a repeat count $R$;\\ an allowed edge set $E$;\\ 
    an $N$ dimensional vector $\vec{h}$ and an upper \\ diagonal $N \times N$ matrix $\hat{K}$ 
    with $h_j, K_{ij} \in {\cal R}$ \\and $K_{ij} \in E$.}
  \Output{An $R$ element array Outcell, for which each element is a set $\{s_j^*\}$, $s_j \in \{-1, +1\}$, which represents a potential minimizer of ${\cal E}(s_1, ..., s_N)=\sum_{j=1}^N h_j s_j + \sum_{i,j \in E} K_{ij} s_i s_j$.}
  \vskip0.1in
  \hrule
  \vskip0.1in
  \set $t_f= 100 \ \mu s$,\ $R$ = 128\;
  \load $\vec{h}$ and $\hat{K}$ values into hardware\;
  \wait 1 $ms$ for hardware to cool down\;
  \For {j = 1 to $R$}{%
  \Indp  Run {\bf annealing  algorithm}\;
    Read out qubits to generate trial solution $\{s_j^*\}$\;
    Set Outcell(j)=$\{s_j^*\}$\;}
\caption{An adiabatic quantum optimization algorithm.}
\label{alg:aqo}
\end{algorithm}

\begin{algorithm}
\SetAlgoLined
\SetKwInOut{Input}{Input}\SetKwInOut{Output}{Output}
\SetKw{set}{set}
\SetKw{wait}{wait}
\Input{A run-time $t_f$;\\ a set of qubits with Hamiltonian $H(s)=A(s) H_I+B(s) H_F$,\\ 
  where $s=t/t_f$, $0 \leq s \leq 1$,\\ $H_I=\sum_{j=1}^N \sigma_{x,j}$,
  $H_F=\sum_{j=1}^N h_j \sigma_{z,j} + \sum_{i,j \in E} K_{ij} \sigma_{z,i} \sigma_{z,j}$,\\
  where $\sigma_{x,j}$ and $\sigma_{z,j}$ \\ are Pauli matrices for qubit $j$, \\ 
  and $A(s)$ and $B(s)$ are envelope functions\\ 
  with units of energy such that $A(0)/B(0) \gg 1$ \\and $A(1)/B(1) \ll 1$.}
\Output{Evolution of $H(s)$ from $H(0)$ to $H(1)$.}
\vskip0.1in
\hrule
\vskip0.1in
\set $s=0$\;
\wait 1 $ms$ for hardware to reach ground state of $H(0)$\;
Ramp currents on global analog lines to drive evolution $s \rightarrow 1$
\caption{The annealing algorithm.}
\label{alg:asa}
\end{algorithm}

Quantum computation intimately ties the physics of the underlying
hardware to its intended algorithmic use. Both the problem to be
solved, and the algorithm used to solve it, are implemented by
manipulating the system Hamiltonian. Primarily motivated by this
observation, the approach we have taken to design hardware is a
top-down one. For the circuits considered here, the requirements are
driven by what is required to run the AQO algorithm. To provide
context for the material in this section we first provide an overview
of the algorithm itself.

Consider the following discrete optimization problem: Given a vector
$\vec{h}$ and upper diagonal matrix $\hat{K}$, where the elements of
both are real numbers, find the set $\{s_i^*\}$ that minimizes the
objective function
\begin{equation}
{\cal E}(s_1, ..., s_N)=\sum_{j=1}^N h_j s_j + \sum_{i,j \in E} K_{ij} s_i s_j \label{obfunc}
\end{equation}
where $s_i=\{-1,+1\}$, and $E$ is an set of $(i,j)$ pairs where
$K_{ij}$ is allowed to be non-zero. We call $E$ the {\em allowed edge
  set}. The necessity for explicitly defining the set $E$ arises
because ultimately we will connect this term in the objective function
to physical couplings between pairs of qubits, and for a variety of
reasons the number of elements in $E$ will generally be much less than
the total number of possible pairs $N(N-1)/2$.  A design constraint on
processor architecture is that qubits must be connected in such a way
so that finding the minimum of Eq. (\ref{obfunc}) is NP-hard. Even
with this constraint, it is straightforward to find realizable sets
$E$ for which this holds, and we will focus exclusively on these cases.

An AQO algorithm exists for solving this problem. The approach is
outlined in Algs. \ref{alg:aqo} and \ref{alg:asa}. The control system
reported on here enters into these in the {\bf load} step of
Alg. \ref{alg:aqo}.

\subsection{Processor Interconnect Architecture}

\begin{figure}[t]
\includegraphics[width=3.3in]{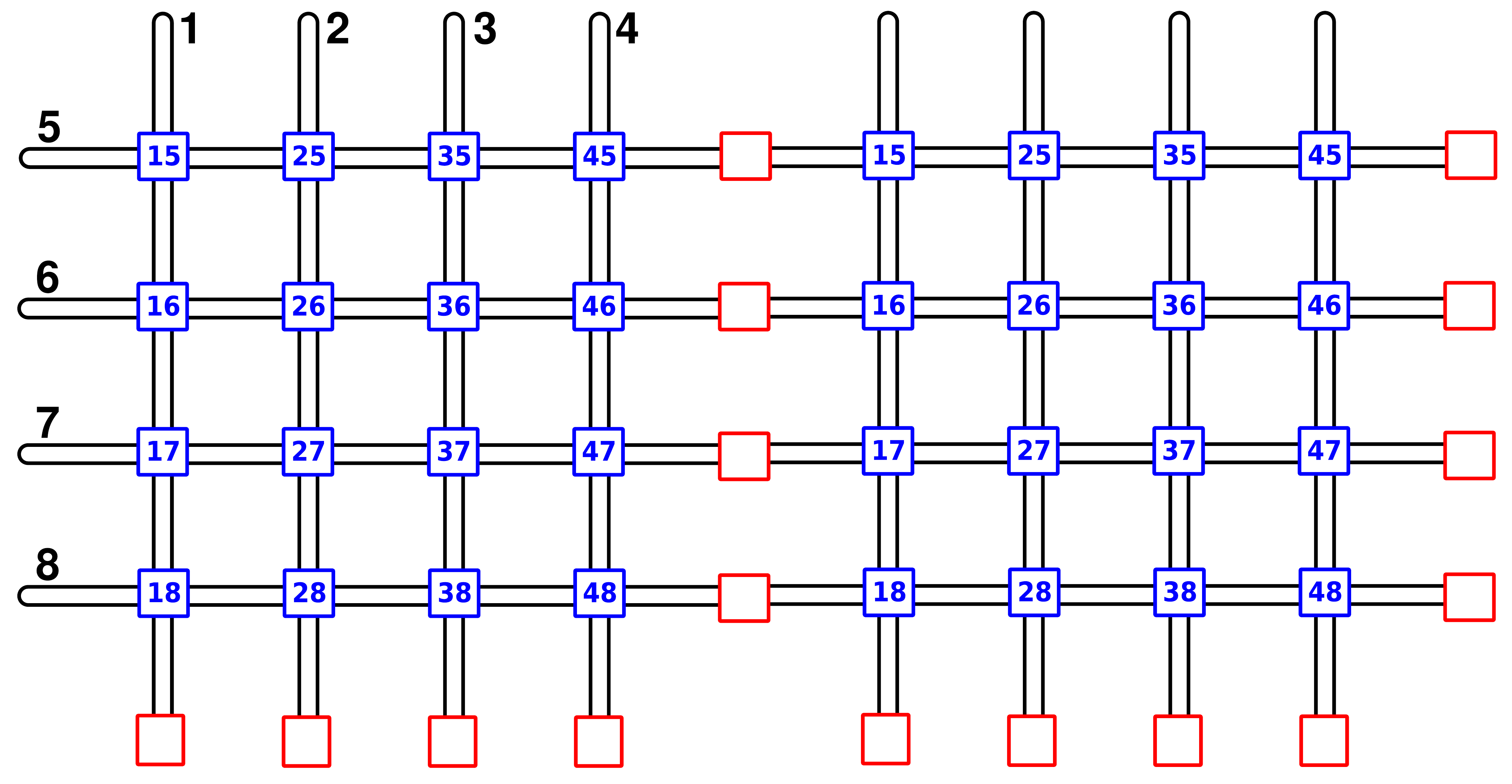} 
\vskip 0.1in
\includegraphics[width=3.3in]{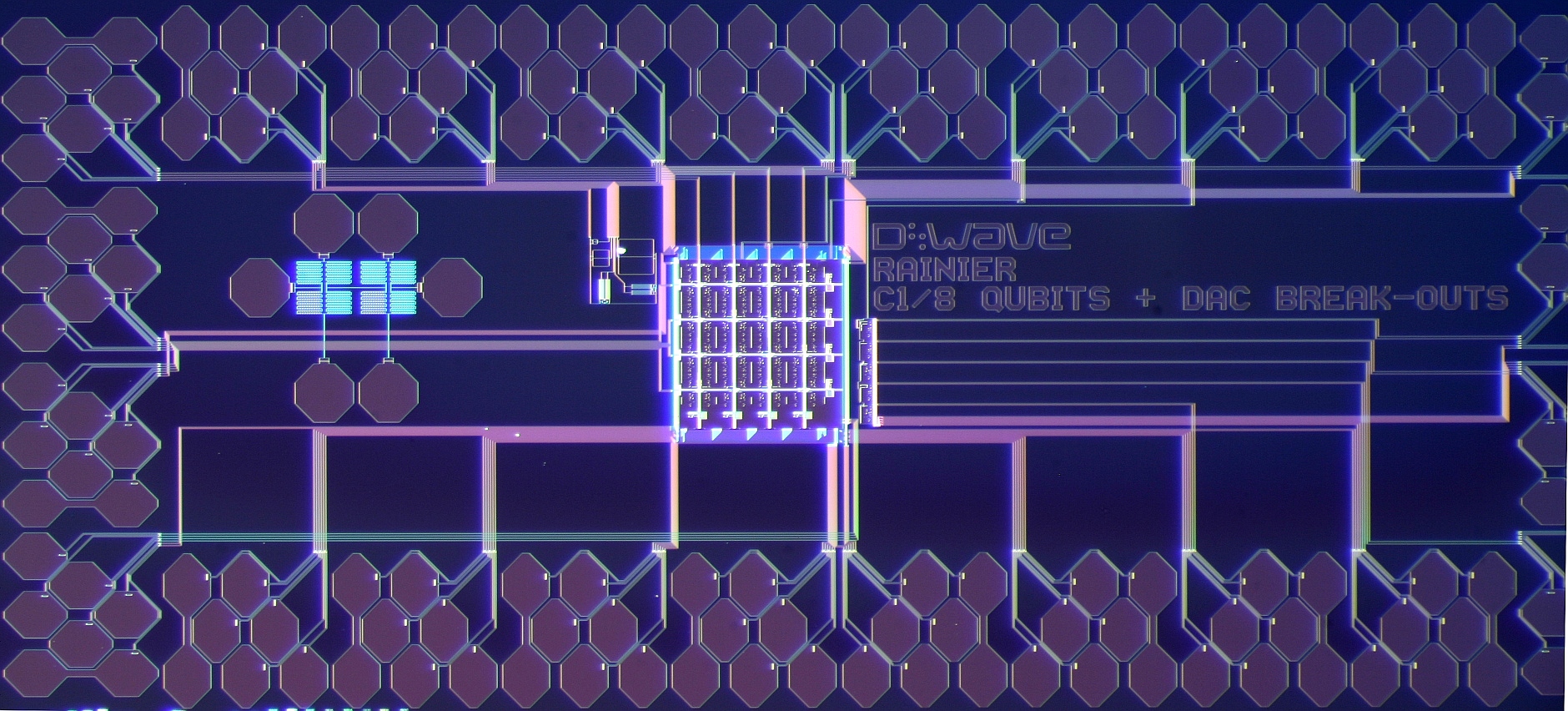} 
\caption{{\bf Top:} Schematic showing two eight-qubit unit cells tiled
  together. The qubits are schematically shown as the extended black
  loops, similar to the way these devices are physically
  implemented. The couplers (shown as blue and red squares) are local
  to the intersections of qubits. {\bf Bottom:} Photograph of a standalone
  eight qubit unit cell occupying a $700\mu\mathrm{m} \times
  700\mu\mathrm{m}$ square on a 3~mm $\times$ 7~mm chip.}
\label{fig:unitcell}
\end{figure}

There are many possibilities for how one might try to build a hardware
system capable of running Algs. \ref{alg:aqo} and \ref{alg:asa}. Here
we focus on a unit cell consisting of eight qubits and 24
couplers. See Fig.~\ref{fig:unitcell} for a schematic and photograph
showing the interconnect pattern. Copies of this unit cell can be
connected together, as indicated in the top of
Fig.~\ref{fig:unitcell}, and this is how we design larger
systems. This choice of unit cell fixes the allowed edge set $E$, and
satisfies the constraint that minimizing Eq. \ref{obfunc} be NP-hard.

\subsection{Number of DACs}

The total number of DACs required for circuits of increasing
complexity is shown in Table \ref{table:scale}.  Here we provide a
brief overview of how these numbers arise, and refer the reader to
\cite{coupler-paper, robust-scalable} for further details.

\begin{table}
\caption{Parts count vs. number of unit cells.\label{table:scale}}
\begin{ruledtabular}
\begin{tabular}{@{}rrrrr}
\noalign{\smallskip}
Unit Cells & Qubits & Couplers & DACS   &   JJs    \\ 
\noalign{\smallskip}\hline\noalign{\smallskip}
1   &   8      & 16       & 56          &   1500   \\ 
4   &   32     & 72       & 232         &   6000   \\
16  &   128    & 328      & 968         &   24000  \\ 
64  &   512    & 1416     & 3976        &   96000  \\
256 &   2048   & 5896     & 16136       &   384000 \\
\noalign{\smallskip}
\end{tabular}
\end{ruledtabular}
\end{table}

\subsubsection{One DAC per Coupler}

A tunable compound Josephson junction rf-SQUID coupler inductively
coupled to qubits $i$ and $j$ is used to set each desired value of
$K_{ij}$\cite{coupler-paper}. One such physical device is required per
element of the allowed edge set $E$. Couplers are controlled using a
static dc flux bias applied to their compound-junction--no time
dependence in this signal is required. For this design, the flux bias
is provided by the DAC shown in red in Fig.~\ref{qubit-fig}.

\subsubsection{Five DACs per Qubit}
\label{sect:qubitdacs}

The potential energy of an ideal compound Josephson junction rf-SQUID
qubit is \cite{han}
\begin{eqnarray}
\mathcal{U} & = & - \frac{\Phi_o \cal{I}_\text{c}}{2 \pi} \cos\left(\frac{2\pi\Phi_q}{\Phi_o}\right)
                        \cos\left(\frac{\pi \Phi_\text{CJJ}}{\Phi_o}\right) \nonumber \\
            & + &  \frac{(\Phi_q-\Phi_q^x)^2}{2\ \lqu}
                  + \frac{(\Phi_\text{CJJ}-\Phi_\text{CJJ}^x)^2}{2\ \lcjj}
        \label{eqn:h2}
\end{eqnarray}
where $\cal{I}_\text{c}$  is the \emph{sum} of Josephson critical currents in the compound
junction, \lqu\ and \lcjj\ are the inductance in the qubit and
compound-junction loop, respectively.  Likewise $\Phi_\text{q}$, $\Phi_\text{q}^x$,
and $\Phi_\text{CJJ}$, $\Phi_\text{CJJ}^x$ are the internal and
applied flux for the qubit and CJJ loop respectively.

\begin{figure}
\includegraphics[width=3.3in]{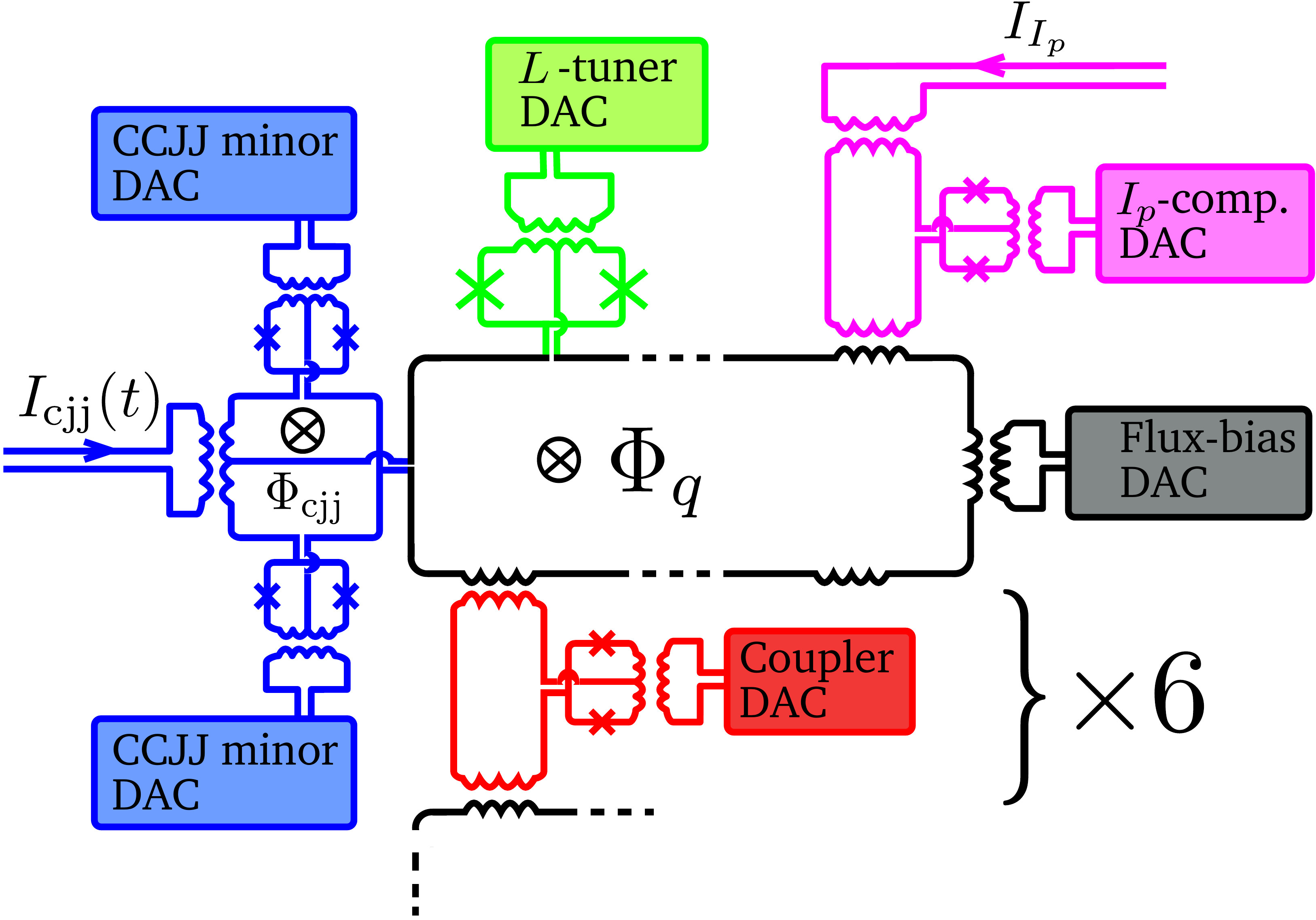}
\caption{Single qubit schematic. The five main parts of this qubit
  design: the qubit main loop (black) with flux-bias DAC providing the
  flux-bias $\Phi_q$, the CCJJ (blue) with cjj-bias $\Phi_{cjj}$ in
  the major lobe and two DACs biasing the minor lobes, the L-Tuner
  (green) with DAC, and \ip-compensator (pink) with DAC. Also shown
  is a coupler (red) with coupler DAC. The two global time dependent
  control lines ($I_{cjj}$ and $I_{Ip}$) used for running the
  annealing algorithm are also shown.}
\label{qubit-fig}
\end{figure}

Eq.~\ref{eqn:h2} is only applicable when the two junctions making up
the compound-junction are identical.  Junction critical current \ic s
of identically drawn Josephson junctions in superconductor fabrication
processes are reported to have a normal distribution with a standard
deviation of anywhere from 1\% to 5\% \cite{abelson95,nakada}.  Thus,
we expect real compound-junctions to be naturally imbalanced.  This
causes difficulties in running the annealing algorithm
Alg. \ref{alg:asa} \cite{robust-scalable}. To overcome the junction
imbalance problem we use a more complex structure which we call a
compound-compound-Josephson junction (CCJJ) which is described in
detail in \cite{robust-scalable}.  This provides two additional degrees of
control freedom per qubit, which can be used to correct for reasonable
junction imbalance ($\sim$ 5\% \ic\ difference). We access these
structures via the blue CCJJ minor DACs in Fig.~\ref{qubit-fig}.

As inter-qubit coupling strength is adjusted, the susceptibility of
the coupler, and the extent to which it inductively loads the qubit,
will change~\cite{coupler-paper}.  This causes the qubit inductance
$L_q$ in Eq.~\ref{eqn:h2} to be dependent upon the choice of $\left\{
  K_{ij} \right\}$.  To overcome the resulting problem-dependent
inter-qubit imbalance, we add an additional compound-junction,
comprising much larger junctions, in series with the qubit
inductance. We call this structure an $L$-tuner~\cite{robust-scalable}.
The Josephson inductance of this compound-junction is modified with
application of a flux bias applied through an on-chip flux DAC, shown
in green in Fig.~\ref{qubit-fig}.

As discussed in \cite{controlled-annealing}, care must be taken during
annealing to ensure that the final Hamiltonian $H_F$, the one encoding
the problem we wish to solve, is that which was intended. Using a
compound-junction to modify the relative weights of $H_I$ and $H_F$
causes $\vec{h}$ and $\hat{K}$ to change during annealing, both in an
absolute sense, and relative to each other. This arises because
although energy scales $\vec{h}$ and $\hat{K}$ are both functions of
the persistent currents in the qubits (\ip), they have different
functional dependencies.  Qubit \ip\ changes during annealing,
distorting $H_F$.

To keep the relative scale constant, the value of the applied flux
used to implement $\vec{h}$ must change during annealing, but
$\Phi_q(t)$ will be different for each qubit, depending on the
intended value of $\vec{h}$ for that problem.  This is accomplished by
giving each qubit another tunable coupler, coupled to both the qubit
and a shared external analog flux bias line. We call this an \ip
-compensator.  Each such coupler is used as a variable gain element,
programmed with its own DAC (the pink DAC in Fig.~\ref{qubit-fig}),
and used to scale a global controlled signal to the locally required
$h_j$.

Finally, each qubit has a DAC that can apply a small dc flux bias to
its main loop (the black DAC in Fig.~\ref{qubit-fig}).

\subsection{Precision and Range Requirements}

\begin{table*}
\caption{Designed flux ranges and minimal flux steps by DAC type \label{table:req}}
\begin{ruledtabular}
\begin{tabular}{cccccc}
\noalign{\smallskip}
  & & & \multicolumn{2}{c}{Max \# $\Phi_0$} & \text{COARSE/FINE}\\ 
\noalign{\smallskip}\cline{4-5} \noalign{\smallskip}
DAC Type   & Span  & min $\Delta \Phi$  & COARSE & FINE & Ratio \\ 
\noalign{\smallskip}\hline\noalign{\smallskip}
Qubit Flux & $25.5~\mathrm{m}\Phi_0 $ & $0.1~\mathrm{m}\Phi_0$      & 17      & 17      & 14.1\\
CJJ Balance   & $66.1~\mathrm{m}\Phi_0$ & $0.4~\mathrm{m}\Phi_0$      & 17      & 17      & 14.1\\
L-Tuner         & $0.465~~\Phi_0$ & $1.1~\mathrm{m}\Phi_0$      & 40      & 10      & 10.7\\
Coupler         & $0.968~~\Phi_0$ & $2.2~\mathrm{m}\Phi_0$      & 40      & 10      & 10.6\\
\noalign{\smallskip}
\end{tabular}
\end{ruledtabular}
\end{table*}

Requirements on precision and range of flux from the DACs ultimately
depend on the precision to which the elements of $\vec{h}$ and
$\hat{K}$ are to be specified.  The system described here was designed
to be able to attain four effective bits of precision on parameters
$h_j$ and $K_{ij}$; in other words, the elements $h_j$ and $K_{ij}$ can
be specified with a relative precision of about 5\%. This does not
mean that the DACs need only four bits of precision. The DAC
requirements are derived from those on Hamiltonian parameters
$\vec{h}$ and $\hat{K}$, based on which aspect of a qubit or coupler is
being controlled.  In our case, requiring four bits of precision in
$\vec{h}$ and $\hat{K}$ typically translates into a requirement of
about eight bits of precision in each of the DACs.

The primary design parameters for each DAC is its dynamic range: how
much flux is it necessary for the DAC to provide, and how
fine a control of that flux is needed.  All of the DACs were designed to
cover their respective ranges in subdivisions of either 300 or 400
steps. However, they differ in the total amount of flux coupled at
maximum range from around 25 $m\Phi_0$ for the qubit flux bias DAC to
as much as 0.9 $\Phi_0$ from the coupler DAC.  A summary of desired
flux ranges and minimal flux steps is shown in Table~\ref{table:req}.

\subsection{Programming Constraints}

In the design discussed here, there are five DACs per qubit and one
per coupler that need to be programmed to implement a specific problem
instance. While the DACs are being programmed, power is applied to the
SFQ circuits in the address tree, and the chip will heat.  The amount
of time we must wait for it to cool afterwards (step 3 in
Alg. \ref{alg:aqo}) depends on the peak temperatures reached by the
various portions of the circuitry, and the relaxation
mechanisms enabling their return to equilibrium~\cite{savin}.
Minimizing overall programming time, including that required to cool,
is an important design constraint, and must be considered when
comparing control circuitry architectures.

Given the block architecture described above, the number of DACs we
must program increases with processor size as shown in
Table~\ref{table:scale}.  While not every DAC will need to be
programmed for each unique configuration of $h_{j}$ and $K_{ij}$ in
practice, in what follows the assumption will be that all are.  In our
multiple flux quantum based encoding scheme, the programming time will
depend on the value programmed.  To estimate the basic scaling with
number of devices, it is probably reasonable to assume that each time
the processor is programmed for a new problem, each DAC stage must
receive on average half of its designed capacity in pulses.  For
example, the coupler DAC would receive about $20 (\mathrm{COARSE}) + 5
(\mathrm{FINE}) = 25 \Phi_o$, the qubit flux bias DAC about $9
(\mathrm{COARSE}) + 9 (\mathrm{FINE}) = 18 \Phi_o$, or roughly 20
$\Phi_o$ per DAC in either case.

Programming speed can then be bought at the expense of additional
input lines and more parallelization - using more, shallower address
trees each with its own separate input.

However, with or without parallelization, one must be able to load all
the pulses without errors.  The more DACs being programmed, the more
pulses there are that must be routed through the address tree with
fidelity, and the smaller the acceptable error probability per pulse.  
For example, per Table~\ref{table:scale}, with $2048$ qubits, we
require 16,136 DACs.  An average problem would require
loading  $\sim 3.2 \times 10^{5}$ flux quanta onto the chip.
If we want 95\% confidence that we can program problems correctly 99
times out of 100, the probability that any flux quanta makes an error
should not be greater than $10^{-9}$.  With 128 qubits, $10^{-8}$ is
sufficient.

Bit error probability in SFQ circuits has been extensively
studied~\cite{herr96,herr99,satchell}.  Satisfying these and even more
demanding error rate requirements at sub-Kelvin temperatures is
straightforward, but needs to be confirmed for any particular
implementation.

\section{Magnetic Memory Architecture}
\label{sect:arch}

\subsection{Two Stage Multiple Flux Quantum DAC with Reset}
\label{sect:dac}

\begin{figure}[h]
\begin{tabular}{c c c}
\includegraphics[height=1.9in]{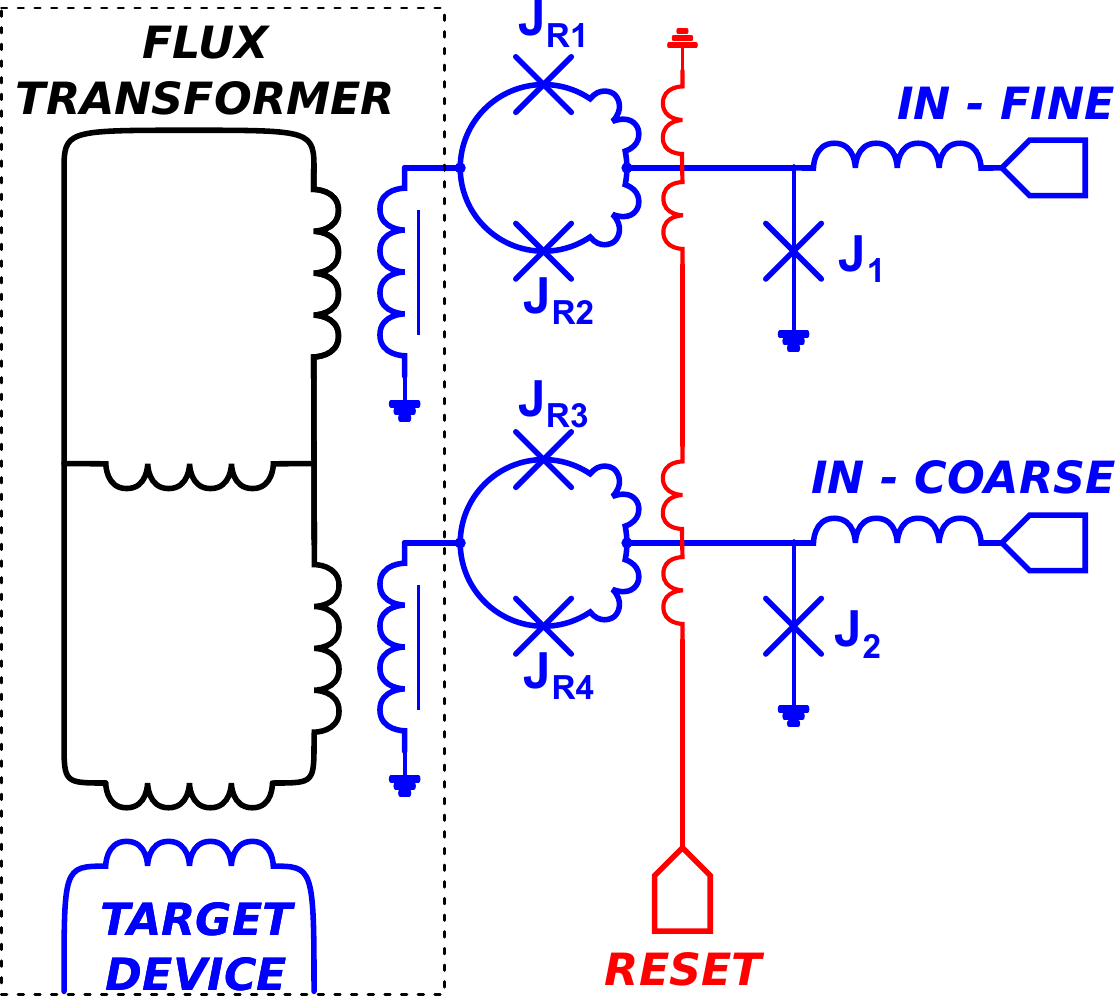} &
\  &
\includegraphics[height=1.9in]{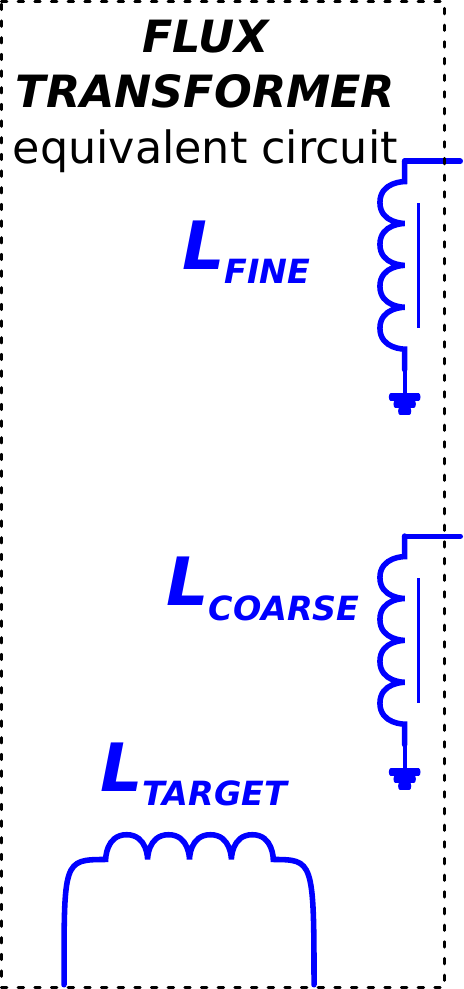}\\
(a) & & (b) \\
\end{tabular}
\caption{(a) Schematic of a two stage DAC made of a pair of separately
  addressable storage inductors (distinguished by an extra line
  alongside the inductor schematic symbol). The flux from each
  inductor is scaled and summed in a flux transformer into the target
  output. Josephson junction critical currents in the schematic are
  $J_1,J_2 = 12.5~\mu A$ and $J_{R1}, ... J_{R4} = 27.1~\mu A$. (b)
  Equivalent circuit for a three-port model of the flux transformer.}
\label{fig:dac}
\end{figure}

The DACs each have two stages.  Each stage comprises a large storage
inductor in series with a two-junction reset SQUID, and an input
junction, as shown in Fig.~\ref{fig:dac}(a).  Each stage was designed
with $\beta \equiv 2 \pi L \ic / \Phi_o$ in the range 75 to 300, 
depending on their function, and thus able to hold in the range of 10
to 40 flux quanta of either polarity.  Here L is the stage inductance,
and \ic\ is the effective critical current of the two junction reset
SQUID.

The two-junction reset SQUID is used to empty the DAC stage of stored
flux.  This is accomplished by applying $\Phi_0/2$ flux to the reset
loop, so that its effective critical current $2\ic^{reset}\cos(\pi
\Phi_x/\Phi_o)$, and thus DAC stage $\beta$, is diminished to below
the level required to store flux. For this reset function to be
effective, it must be possible to suppress the effective critical
current of the reset SQUID to less than that required to store one
$\Phi_o$ in the storage inductor.  This requirement places an upper
bound on the DAC stage $\beta$.  It also places a requirement on how
closely matched the \ic s of the two junctions in the reset SQUID must
be to each other.  This is because the minimum effective critical
current will not be less than the difference in the \ic s of the two
reset junctions. Thus given a particular fabrication process, with its
feature size, penetration depths, and characteristic junction \ic\ 
spread, there will be some maximum number of $\Phi_o$ that can be
stored in a DAC that can be reliably reset to an empty state. This
limits the dynamic range of an individual DAC stage.

\begin{table}[h]
\caption{ The three-port inductance
  matrix for the flux transformer shown in Figure~\ref{fig:dac}(b) was extracted
  from the layout using {\tt FastHenry 3.0wr}~\cite{wrcad}}
\begin{ruledtabular}
\begin{tabular}{lddd}
\noalign{\smallskip}
       & \text{Fine}    & \text{Coarse}   & \text{Target}       \\ 
\noalign{\smallskip}\hline\noalign{\smallskip}
Fine   & 3.50~nH & 9.9~pH &  0.7~pH   \\
Coarse &         & 3.90~nH  & -10.2~pH     \\
Target &         &          &  7.8~pH \\
\noalign{\smallskip}
\end{tabular}
\end{ruledtabular}

\label{tab:3port}
\end{table}

We can achieve a dynamic range greater than that of an individual DAC
stage, and shorten programming times, by connecting two or more stages
together, as indicated in Fig.~\ref{fig:dac}(a). The intervening
transformer couples the different stages into the target circuit with
different weights.  The flux transformer can be thought of as playing
the role of an R-2R ladder, such as is frequently used in construction
of semiconductor DACs.  One important difference is that successive
stages of this DAC differ from each other not by factors of two, but
more typically by a factor of ten, depending on the requirements set
by the target device.  The other difference is that we are
transforming and dividing flux rather than voltage.

The flux transformer can be modeled with the equivalent circuit shown
in Fig.~\ref{fig:dac}(b), where we consider only two storage inductors
$L_{COARSE}$, $L_{FINE}$, and the target inductor $L_{TARGET}$.  The
inductance matrix for the equivalent circuit shown in
Fig.~\ref{fig:dac}(b) was calculated by modeling the layout using {\tt
  FastHenry 3.0wr}~\cite{wrcad}, and is shown in Table~\ref{tab:3port}.

The flux coupled out of this DAC can be summarized by the expression:
\begin{equation}
\label{eq:dac}
 \Phi_\mathrm{OUT} = k \cdot \left(\mathrm{N_{COARSE}} \Phi_0 
 + \frac{1}{\gamma} \cdot \mathrm{N_{FINE}} \Phi_0 \right)
\end{equation}
where $N_\mathrm{COARSE,FINE}$ represent the integer number of flux
quanta that are stored within the respective DAC stages, $k$ the
coupling constant describing the amount of flux from the COARSE stage
into the output device, and $\gamma$, the division ratio between
COARSE and FINE, which is typically 10 for the devices discussed
here.

\subsection{DAC Noise}

One concern with using SFQ circuitry in this way arises from the fact
that its Josephson Junctions are usually critically damped with
external shunt resistors.  These resistors are a source of
fluctuations which may ultimately decrease the precision with which
Hamiltonian parameters $h$ and $K$ can be specified.  In the design
presented in Figure~\ref{fig:dac}(a), a number of factors serve to
isolate the resistors in the SFQ circuitry from the junctions in the
qubit, so that their impact on the qubit can be quite small.

To see this more clearly, we consider a simple lumped element circuit
model of the DAC and qubit.  A description of our system using lumped
elements is reasonable at low enough frequencies, but will eventually
fail at higher frequencies, for example when the 1~millimeter long
coils in the DAC spirals approach $\lambda/4$ at around 100 gigahertz.
Fortunately, we are most concerned with fluctuations occuring at an
energy scale comparable to, or less than, the tunnel splitting of our
qubits during the annealing algorithm, typically a few gigahertz or
less\cite{robust-scalable}.

\begin{figure}
\includegraphics[width=2.75in]{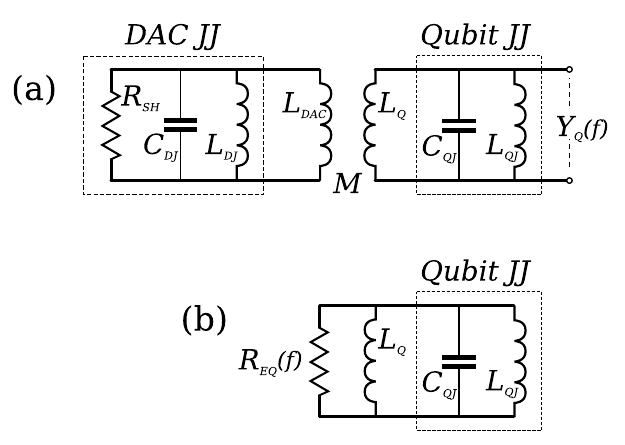}
\includegraphics[width=3.0in]{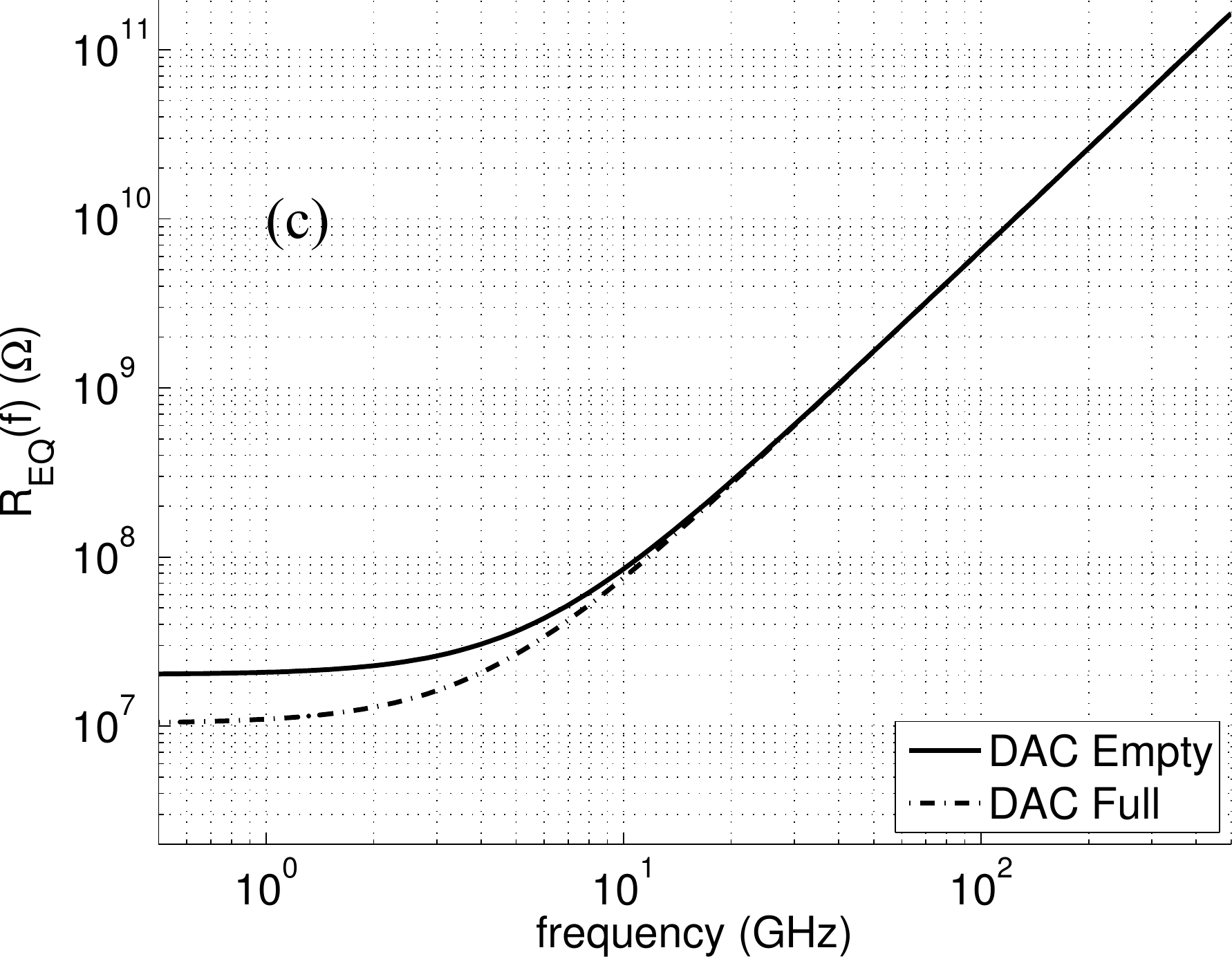}
\caption{
A simplified equivalent circuit of (a) DAC storage inductor coupled to 
qubit, (b) a simpler equivalent circuit for comparison. (c) a plot of 
equivalent resistance for $R_{EQ}(f)$ for the case of DAC empty(full) 
for parameters $R_{SH} = 0.9 \Omega$, $L_{DJ} = 26~pH (36~pH)$, 
$C_{DJ} = 160 fF$, $L_{DAC} = 3.6~nH$, $M = 10.2~pH$, $L_Q = 320 pH$, 
$C_{QJ} = 200~fF$, $L_{QJ} = 91~pH$.
\label{fig:dacnoise}}
\end{figure}

There are many shunted junctions in the SFQ circuits discussed here,
but none couples more strongly into the qubit than the input junctions
of the various COARSE DAC stages, labelled $J_2$ in
Figure~\ref{fig:dac}(a).  In what follows, the effect of this
junction's shunt resistor on a qubit is considered by analyzing an
equivalent circuit shown in Figure~\ref{fig:dacnoise}(a) for the case
of the qubit flux bias DAC. Here the DAC input junction has been
linearized, and is represented by its Josephson inductance $L_{DJ}$
and junction capacitance $C_{DJ}$. Two different operating points of
the DAC input junction, corresponding to the DAC being empty or full
of flux, are used to determine the small signal inductance
$L_{DJ}$. The qubit's four CCJJ junctions are represented as a single
linearized junction described by $L_{QJ}$ and $C_{QJ}$.

One way to characterize the extent to which fluctuations in $R_{SH}$
couple into qubit junctions is by comparing its effect to that of an
equivalent shunt resistor $R_{EQ}$ connected directly across the qubit
junctions, as indicated in Figure~\ref{fig:dacnoise}(b).  The
magnitude of $R_{EQ}$ will be frequency dependent, and can be
determined as $R_{eq}(f) = (\Re{Y_Q(f)})^{-1}$ where $Y_Q$ is the
admittance of the circuit across the terminals shown in
Figure~\ref{fig:dacnoise}(a).  The resulting $R_{EQ}$ is plotted
vs. frequency in Figure~\ref{fig:dacnoise}(c) for circuit parameters
described in the caption. At low frequency $R_{EQ}$ is around
$10~M\Omega$ and grows at frequencies above a few gigahertz.

Even at low frequency, most current fluctuations from $R_{SH}$ are
shunted by the DAC input junction, whose small-signal inductance is
between 100 and 150 times smaller than the DAC storage inductor
$L_{DAC}$, depending on the state of the DAC.  Noise current that does
flow into $L_{DAC}$ can in turn couple into the qubit, though it is
further reduced by a factor of $M/(L_Q+L_{QJ})$, or around 1/40 for
the circuits discussed here.  At higher frequencies fluctuations are
further shunted by the junction capacitances.

The other DACs couple into the qubit in a less straightforward
fashion, but we can still make a reasonable estimate of their impact.
An example is the case of the L-Tuner DAC.  The L-Tuner is a dc-SQUID
connected in series with the qubit inductor\cite{robust-scalable}.  It
is flux biased in such a manner that it would not apply any flux into
the qubit if its two $8.5~\mu A$, $1.85~\mu m$ diameter junctions were
identical, but does so when they differ.  A 1\% mismatch between
junction critical currents is typical for junctions of this size in
our process.  A typical operating point for the L-Tuner corresponds to
a flux bias of $\Phi_0/4$. For this case, following Eq.~4c from
\cite{robust-scalable}, about 1\% of the flux applied to the L-Tuner
will get applied into the qubit body.  This is comparable but smaller
than the corresponding factor of $M/(L_Q+L_{QJ})$ of 1/40 discussed
for the qubit flux bias DAC.  For this reason, we expect the
corresponding $R_{EQ}$ for the case of the L-Tuner DAC to be larger
than that of the qubit flux bias DAC.  It should then have a
relatively smaller effect on the qubit. Similar arguments apply to the
other DAC types.

An $R_{EQ}$ of $10~M\Omega$ will contribute approximately $L_Q\sqrt{4
  k_B T / R} \simeq 50 p\Phi_0/\sqrt{Hz}$ flux noise into the qubit,
considerably less than the 1.3~$\mu \Phi_0/\sqrt{Hz}$ at 1 Hz observed
in 1/f noise in our qubits \cite{robust-scalable,trevornoise}.  Thus
we do not expect the shunt resistors in our control circuitry will add
a significant amount of flux noise to our qubits.

\subsection{Programmable time-dependent signals}
\label{sect:ipcomp}

As discussed in section~\ref{sect:qubitdacs}, we require the ability
to supply time dependent signals to each of the qubits to compensate
for the fact that the qubit persistent current changes during the
annealing process.  These time dependent signals need to have the same
temporal shape but with different magnitudes.  The DACs discussed
above can hold static flux, and are not suited to provide real-time
signals.  This is because they do not include a sample-and-hold stage
to protect the output from transients during programming.  Moreover,
real-time updating of the DACs would raise the temperature of the chip
to an unacceptable level.

\begin{figure}
\includegraphics[width=3.3in]{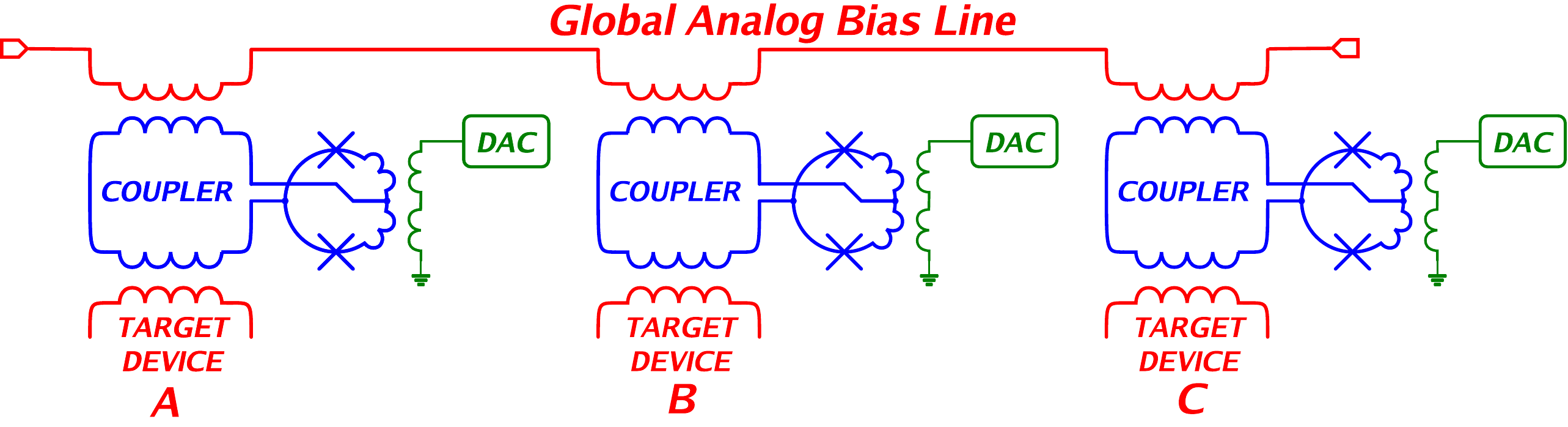}
\caption{A time dependent current on a global analog bias line can be
uniquely scaled into each of several target devices by using
independent programmable gain elements (blue), each controlled with its
own DAC (green).}
\label{fig:ipcomp}
\end{figure}

Rather, time dependent signals can be customized using the tunable
coupler discussed in Reference~\cite{coupler-paper} as a variable or
programmable gain element.  A global analog bias line holding a
\emph{master copy} of the desired time dependent signal is coupled to
each qubit on the chip through its own programmable gain element, as
indicated in Fig.~\ref{fig:ipcomp}.  Each programmable gain element is
controlled by its own DAC.  In conjunction with an additional DAC (not
shown) to provide a flux offset to each target device, the
\emph{master copy} of the signal can be uniquely transformed for each
target in the following manner:
\begin{equation}
\label{eq:ipcomp}
    \Phi_i(t) = a_i + g_i \Phi_{global}(t)
\end{equation}
where $a_i$ and $g_i$ are programmable on a per device basis.  This is
not as flexible as having independent arbitrary waveform generators
for each device, but it is flexible enough to satisfy the requirements
of the $I_P$ compensator.

\subsection {Demultiplexer Tree}
\label{sect:demux}

\begin{figure}
\includegraphics[width=3.3in]{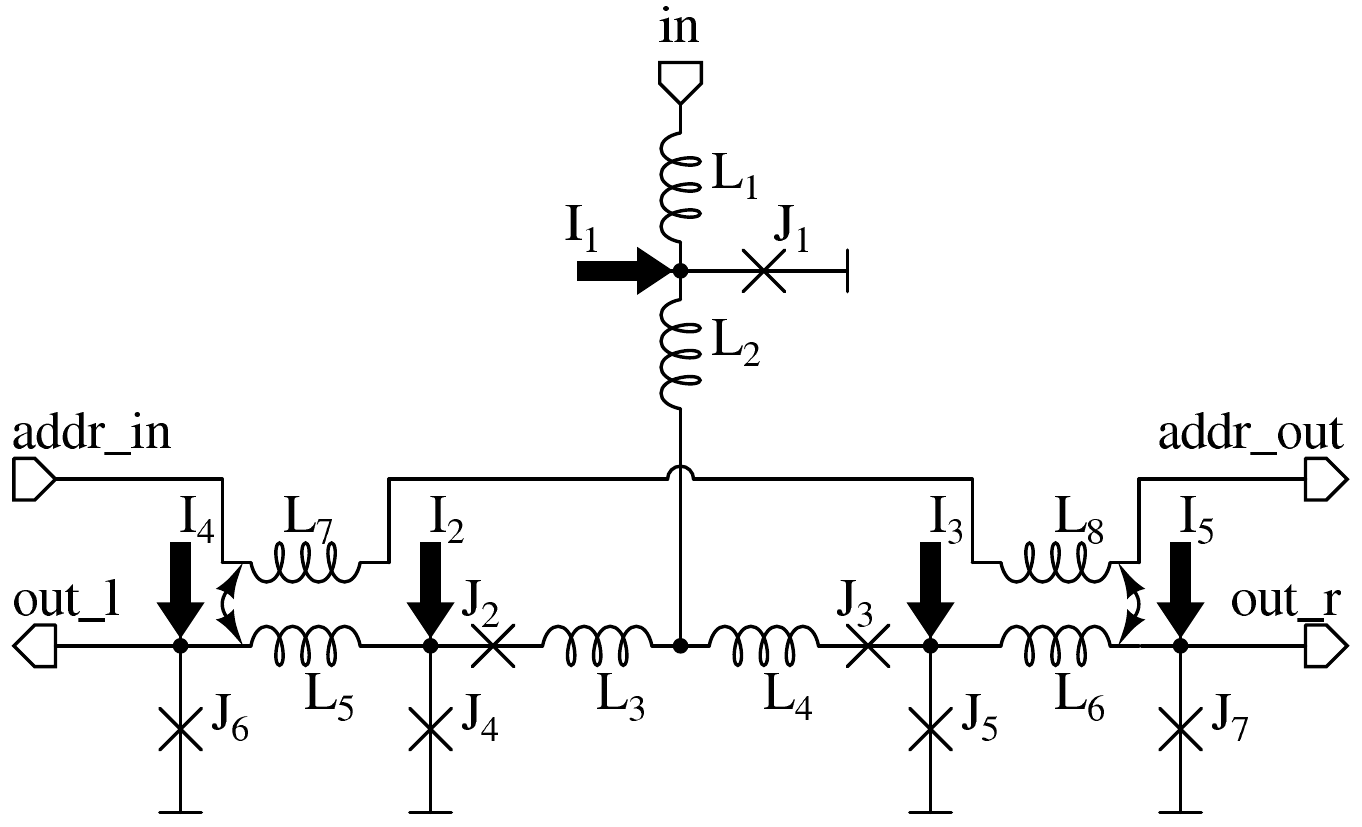}
\caption{1:2 demultiplexer gate used to construct the address tree.
All junctions are explicitly shunted with TiPt resistors (not shown in
schematic). Large arrows represent bias current that is supplied
through bias resistors from a common voltage rail. 
In the schematic, J1,J4-J7 $= 10.6~\mu A$, J2,J3 $= 8~\mu A$, L1 = 26~pH, L2 = 41.6~pH,
L3,L4 = 12.2~pH, L5,L6 = 42.5~pH, both M's = 2.3~pH,
I1 $= 4.4~\mu A$, I2,I3 $= 6.6 \mu A$, and I4,I5 $= 7.8 \mu A$
}
\label{fig:demux}
\end{figure}

The DACs discussed above were loaded with SFQ pulses routed through a
binary tree demultiplexer circuit shown in Fig.~\ref{fig:tree}.  Each
address tree is fed SFQ pulses originating in an SFQ generator
circuit, namely a flux biased dc-SQUID.  Each node of the tree is made
of a 1:2 SFQ demultiplexer circuit, as shown in Fig.~\ref{fig:demux}.
The 1:2 demultiplexer circuit is addressed with a magnetically coupled
flux bias line which steers an incoming SFQ pulse to one of its
outputs based on the sign and magnitude of current on that address
line.

Reversing the polarity of the bias current allows flux quanta of
opposite polarity to be routed, though the sign of the address current
must also be reversed to get this negative flux to the same output
port.  This makes use of a symmetry not commonly exploited in RSFQ
circuits.  Here it allows the DAC stages to store flux of either sign,
and allows the state of each stage to be both incremented and
decremented.  This in turn allows DAC programming to be performed
incrementally - starting from the previously programmed state without
first resetting to the empty (no stored flux) state.  

Address lines are shared for all demultiplexer nodes at a particular
depth of the tree.  The final address line in the tree chooses between
COARSE and FINE stages of each DAC.  The 64 DACs mentioned above are served
by two separate trees, each addressing 32 DACs.  These trees require
five address lines to address a particular DAC, plus a sixth line to
choose between FINE and COARSE, giving a total of six address lines to
service the circuit block.

\section{Demonstration of control circuit functionality}
\label{sect:data}
For the circuit block discussed in this paper, the control circuitry
in its entirety represents moderate complexity - certainly not the
most complex or heterogeneous SFQ circuit demonstrated to date, nor
the one with the most junctions. The eight qubit circuit block
reported here, including the attached control circuitry, contains just
over 1,500 Josephson junctions and 2,000 resistors.  Nevertheless,
implementation of a new design in a new foundry requires careful
performance evaluation.  We must determine that the circuit yielded,
operated as designed, and whether variances are due to design or
fabrication issues.  We must determine if it meets its design
requirements.

\begin{figure}
\begin{center}
\includegraphics[width=3in]{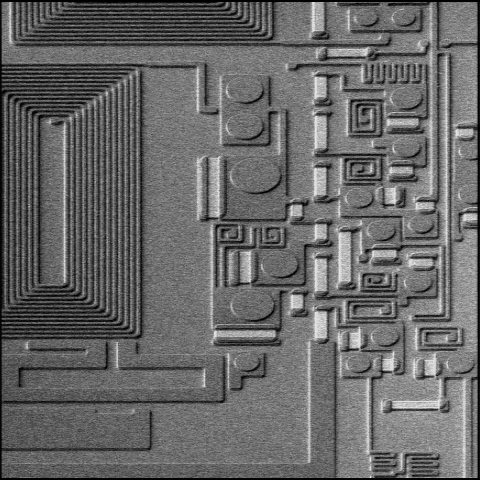}
\end{center}
\caption{SEM image of a portion of the DAC and demultiplexer circuitry
after deposition and patterning of the resistor layer and the trilayer
steps, but prior to applying the upper three dielectric and metal
layers.}
\label{fig:sem}
\end{figure}

A scanning electron microscope (SEM) image of a portion of one of the
DACs and demultiplexer cells equivalent to those reported here is
shown in Fig.~\ref{fig:sem}. The image was taken after patterning of
the trilayer and subsequent junction definition.  Uncontacted
junctions appear as circles and are still visible in the demultiplexer
circuitry toward the right of the image. Subsequent fabrication involves
the deposition of three planarized dielectric layers and three additional
metal (Nb) layers. Junctions in this circuit have critical currents in
the range $10 - 20 \mu\text{A}$. Resistors are also visible and appear as
rectangles contacted at each end. Bias resistors are long and thin
while shunt resistors for this portion of the circuitry are relatively
wide. The coils visible on the left are portions of the two storage
inductors for one of the L-Tuner DACs.  The coils are patterned with
$0.25\mu\text{m}$ lines and spaces.  The DAC's reset junctions are visible in
the upper center of the image.  To the right is one of the demultiplexer
cells.  The entire field of view is approximately $40~\mu\text{m}$ in width
and $50 \mu\text{m}$ top to bottom.

In the architecture described above, many of the DACs are embedded
deeply within the circuit, with no convenient or direct method to
determine how much flux they actually apply to the target per $\Phi_o$
in the FINE and COARSE stages.  This inconvenience is addressed in two
ways: Each variant of DAC is implemented in a separate stand-alone or
\emph{break-out} circuit in which it applies a flux bias directly to a
two-junction dc-SQUID.  The dc-SQUID $I_c$ vs $\Phi$ modulation curve
is then measured vs. DAC state, and a precise calibration of FINE and
COARSE weights ($k$ and $\gamma$) can be extracted. Data from
parameters extracted in this way is presented in
Table~\ref{table:achieved}. Second, within the body of the circuit
block, wherever a DAC is used to apply a flux bias, an analog line is
also used to flux bias that target device in parallel with that
DAC. This combination is indicated in the inset of
Fig.~\ref{fig:cplrodac}.  This single analog line is shared amongst
all like control nodes for all qubits, so that only a handful of such
lines are required to service the entire chip. Of course, the shared
analog line cannot be used for independent control of all devices
simultaneously, but it is nevertheless useful for testing individual
devices. For example, each qubit is flux biased by its own DAC
\emph{and} a single shared externally accessible analog line.  The
qubit degeneracy point is easily measured~\cite{readout}.  If the DAC
is then programmed with, for example, $+5\Phi_o$ in its COARSE stage,
one can determine the change in current on that analog qubit flux bias
line required to compensate for the shift in degeneracy point. This
allows us to find the ratio of mutual inductance between the analog
line and the qubit to that between each DAC stage (COARSE \& FINE) and
the qubit.  We can independently measure the mutual inductance of the
analog line into the qubit by noting the $\Phi_o$ periodicity in its
response.  We can then determine $k$ and $\gamma$ for that DAC, which
are the parameters we need to determine how much flux the DAC applies
to its target.

This feedback measurement is applicable to determining $k$ and
$\gamma$ for DACs used for various types of control, not just qubit
flux bias.  The only thing that differs is the nature of the measured
quantity.  For the qubit flux bias DAC, the qubit degeneracy point is
used.  For the L-Tuner DAC, a measure of the qubit's inductance,
ultimately a measure of its circulating current, can be used.  For the
CCJJ DACs, a measure that quantifies the imbalance in the qubit's
compound-junction is used. In all cases we use the measured quantity
to determine what analog signal is necessary to compensate for a
change in the programmed DAC state.  In this way we can determine how
much flux each DAC on chip applies to its target.

\subsection{DAC Biasing a dc-SQUID}
\label{sect:dac-dc-squid}
\begin{figure}[t]
\begin{center}
\includegraphics[width=3.3in]{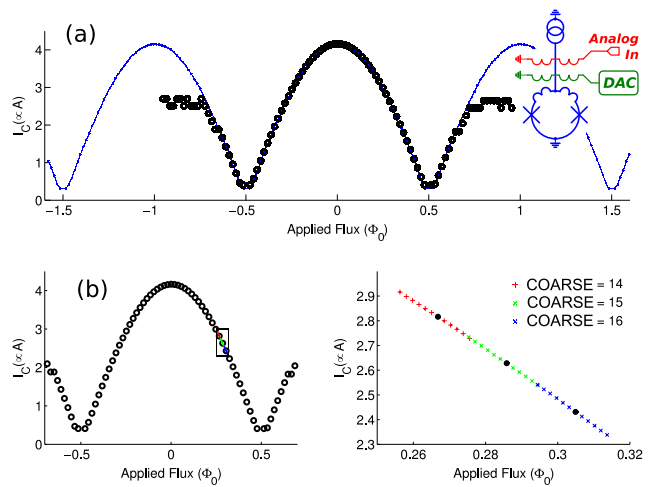}
\caption{ (a) $I_c$ vs. applied flux modulation curve of a small
  $\beta$ two junction hysteretic SQUID where flux was applied with
  (blue dots) external analog bias current and (black circles) a
  two-stage superconducting DAC with greater than $\Phi_0$ total span.
  The black circles are plotted for integer units of $\Phi_0$ sent to
  the COARSE DAC stage. (b) SQUID modulation curve with integral COARSE
  values along with that taken exercising the FINE DAC stage around COARSE
  = 14 (red), 15 (green), and 16 (blue).  (c) Expansion of boxed
  region in (b). Schematic shown in inset. }
\label{fig:cplrodac}
\end{center}
\end{figure}

The couplers shown in red in Fig.~\ref{fig:unitcell} are designed to
couple qubits between different unit cells. For the eight qubit unit
cell under study here, these were not connected to qubits on both ends.
Instead, the inter-unit cell couplers were wired up in such a way that
their compound-junction, still biased by its own DAC, could be
operated as a hysteretic dc-SQUID.  This coupler's DAC could thus be
used to apply flux to a dc-SQUID, and so we traced out the \ic\ vs
$\Phi$ threshold characteristics for this dc-SQUID as discussed in
Reference~\cite{readout}. The \ic\ vs $\Phi$ curve shown in
Fig.~\ref{fig:cplrodac} was taken both with an analog flux bias controlled
directly from room temperature electronics ((a), blue dots) or using
the DAC ((a), black circles).  To make such a plot, we have to know
the coupling constant $k$ between the COARSE DAC stage and the
dc-SQUID it biases, as well as the mutual inductance between the
analog bias line and the dc-SQUID.  As mentioned earlier, the mutual
inductance between analog line and dc-SQUID is easily determined by
observing the periodicity of the modulation curve.  Also discussed
earlier, the coupling constant $k$ and division ratio $\gamma$ are
measured separately with a feedback procedure.

The two threshold curves shown in Fig.~\ref{fig:cplrodac}a begin
to deviate from each other past about $\pm 0.65 \Phi_0$.  This
corresponds to where the COARSE stage of this DAC has reached its
capacity, and fails to store additional flux.  To be clear,
the black circles are plotted vs. {\em flux programmed into} the DAC 
COARSE stage, not flux actually applied by that DAC stage to the dc-SQUID.  
A plot of the latter would fall on top of the blue dots.

Fig.~\ref{fig:cplrodac}b shows the same threshold curve vs.  flux
programmed into the COARSE stage (black dots), but in addition, at
COARSE values of $+14 \Phi_0$, $+15 \Phi_0$, and $+16 \Phi_0$, flux
ranging from $- 6 \Phi_0$ to $+ 6 \Phi_0$ is programmed into the FINE
DAC stage as well.  This is shown in the boxed region in
Fig.~\ref{fig:cplrodac}b, which in turn is expanded in
Fig.~\ref{fig:cplrodac}c.  In Fig.~\ref{fig:cplrodac}c, it is clear that
there is sufficient range in the FINE DAC stage to bridge the COARSE
DAC steps. 

It is worth noticing that the range of the FINE DAC achieved from
adjacent COARSE settings overlap each other. It is essential that they
do not {\emph underlap}, as this would result in holes in the range of
flux that could be provided by the DAC. This means that there can be
more than one way to obtain a particular output FLUX.  But while the
ranges overlap, the specific levels achieved from adjacent COARSE
settings do not, in general, line up.  Of course there is no need for
them to.

To use the DAC, it is necessary for the reset function to operate.  A
reset operation is made by first increasing the flux bias on the reset
SQUID (shown in Fig.~\ref{fig:dac}) to a predetermined level, and then
lowering it back to zero.  The predetermined level is chosen to be
just beyond the current corresponding to a flux bias of $\Phi_o/2$ on
the reset SQUID.  This reset pulse was usually adequate to reset the
DAC. There were cases when the \ic s of reset junctions for a
particular DAC stage differed from each other by more than about 5\%,
where a single reset pulse was not sufficient, and the DAC could
retain one or two $\Phi_o$ after the reset. In these cases, the reset
pulse had to be repeated several times to consistently empty the DAC.
While the correlation between junction spread, DAC stage $\beta$, and
reset function is not quantitatively understood by us, we expect the
problem to worsen with increased junction \ic\ spread, and with
increased DAC stage storage capacity ($\beta$). We found that for the
chips reported upon here, it was always possible to reset all of the
DAC stages by applying multiple reset pulses.

\subsection{DAC applying flux bias to a qubit}

\begin{figure}
\begin{center}
\includegraphics[width=3.3in]{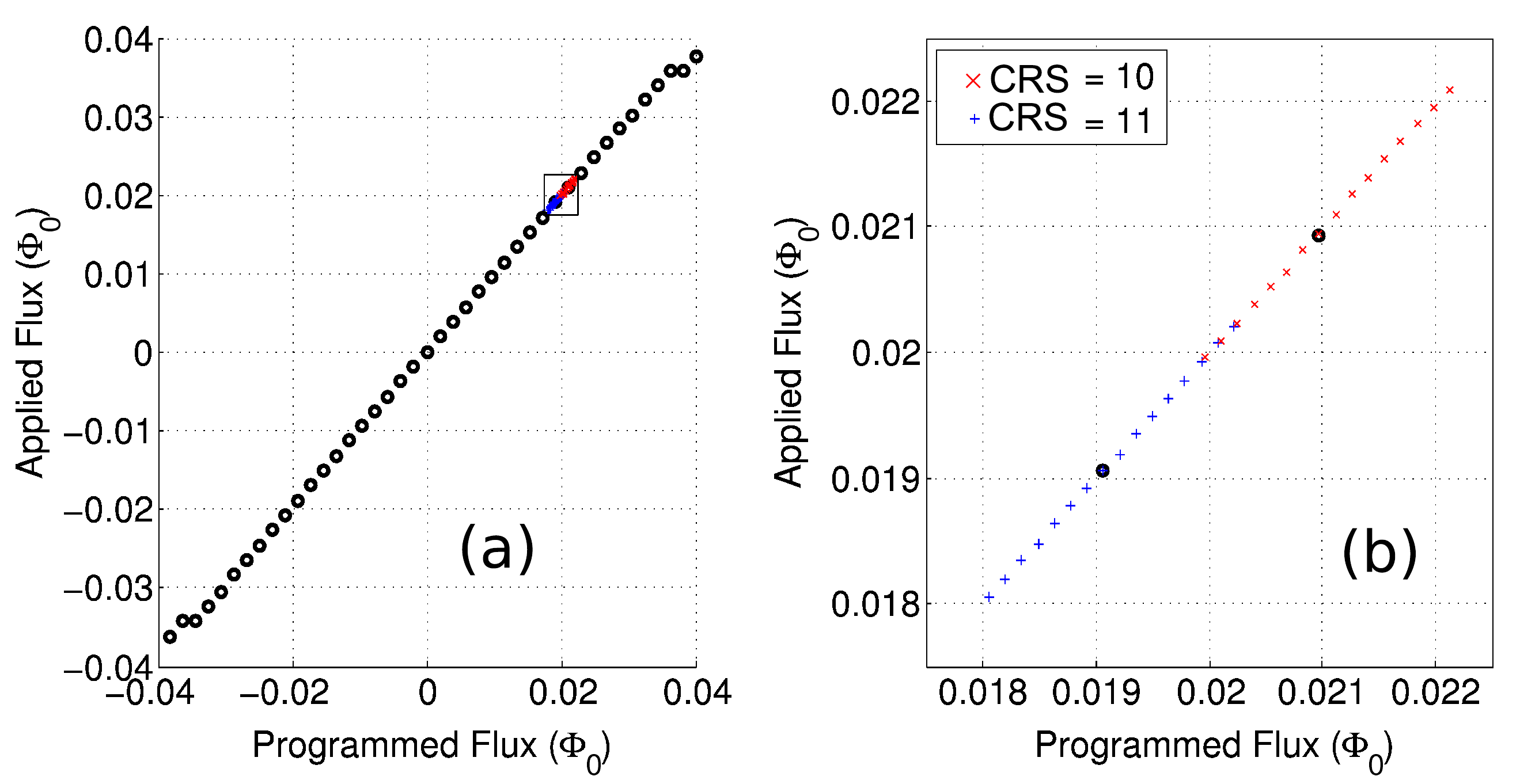}
\caption{(a) Flux response vs. flux programed into the COARSE (black
  circles) qubit flux DAC. (b) Expanded scale of the boxed region in
  (a) shows response of the FINE DAC stage to flux exercised around
  COARSE = $10 \Phi_o$ (blue +) and $11 \Phi_o$ (red x).  Measurement
  uncertainty in flux is smaller than the plot symbols.}
\label{fig:qfbdac}
\end{center}
\end{figure}

Each qubit has a DAC that can apply a flux bias to its body.
Fig.~\ref{fig:qfbdac} shows the flux response relative to the COARSE
and FINE stages for one of these DACs. The limits of the COARSE stage
capacity are just visible at the extrema of the plot. To adequately
cover the range, the maximum span achievable with the FINE stage must
be enough to cover one step of the COARSE stage. This coverage is
clearly attained in Fig.~\ref{fig:qfbdac}.

\subsection{DAC control of inter-qubit coupling}

\begin{figure}
\begin{center}
\includegraphics[width=3.3in]{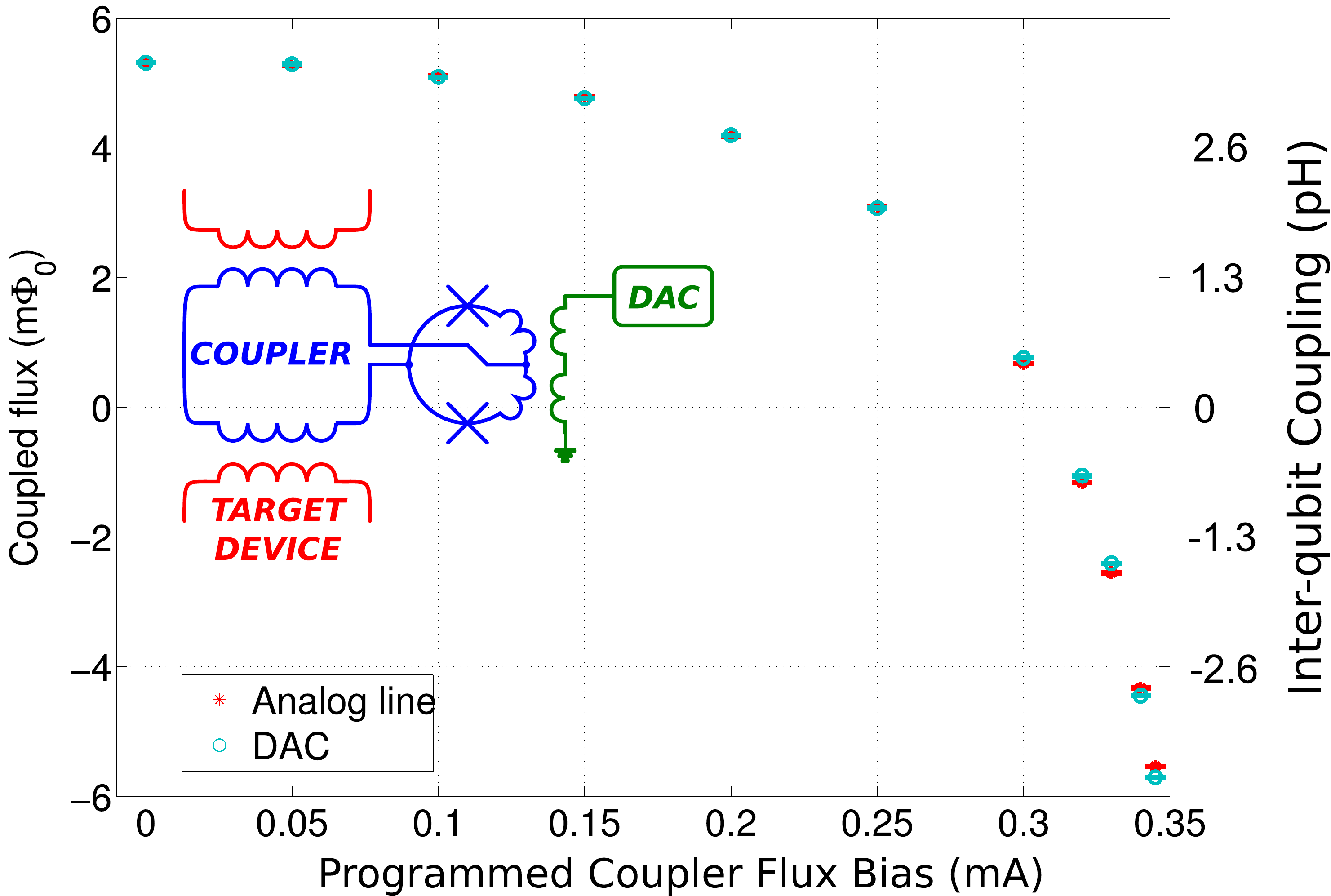}
\end{center}
\caption{Effective mutual inductance of an inter-qubit coupler as a
  function of flux applied by an analog control line (red).  The same
  vs. flux applied by the coupler DAC are shown in blue.  Error bars
  on the plot are smaller than the symbols.}
\label{fig:coupler}
\end{figure}

One of the most challenging cases to treat in the design of these DACs
was the situation in which a large flux span needed to be applied to a
low inductance SQUID.  This situation is most extreme in the case
of the coupler DAC and the $L$-tuner DAC.  In fact, the dc-SQUID
threshold curve presented in Fig.~\ref{fig:cplrodac} above is an
example of such a case - these dc-SQUIDs are patterned identically to
the compound-junction used in the coupler and $I_p$-compensator.

We can also observe a DAC controlling inter-qubit coupling.
Fig.~\ref{fig:coupler} shows a plot of the effective coupling between
two qubits, via a tunable coupler, as a function of flux applied to
control the coupler by an analog control line (red points).
Inter-qubit coupling using the DAC to control the same coupler
is shown as blue circles in the same plot.  More details about the
type of measurement used to obtain this plot are discussed in
\cite{coupler-paper}.

While presented as an inter-qubit coupler, a similar device is
employed in customizing shared time-dependent signals, as discussed in
section~\ref{sect:ipcomp}.  The key difference is that in
Fig.~\ref{fig:ipcomp}, one of the qubit ports on the coupler is
connected to a global analog bias line.  We found it straightforward
to operating this programmable coupler in the fashion described in
section~\ref{sect:ipcomp}.

\subsection{Summary of DAC performance}

\captionsetup[subtable]{position=top}
\begin{table*}
  \caption{Designed vs. achieved DAC parameters}\label{table:achieved}
  \subfloat[COARSE and FINE DAC step weights]{
  \begin{ruledtabular}  
  \begin{tabular}{ldddd}
    \noalign{\smallskip}
    & \multicolumn{2}{c}{COARSE Step $(\mathrm{m} \Phi_0)$} 
    & \multicolumn{2}{c}{FINE Step $(\mathrm{m} \Phi_0)$}  \\ 
    \noalign{\smallskip}\cline{2-5}\noalign{\smallskip}
    DAC Type        & \text{Design} & \text{Achieved} & \text{Design} & \text{Achieved}  \\  
    \noalign{\smallskip}\hline\noalign{\smallskip}
    Qubit Flux      & 3.0  &  3.506(3) & 0.21   & 0.268(3)  \\ 
    CCJJ            & 5.6  &  3.899(2) & 0.40   & 0.296(3)  \\
    L-Tuner         & 11.3   &  8.481(1) & 1.1    & 1.061(1)  \\ 
    Coupler         & 23.6   & 19.0221(2) & 2.2    & 1.788(1) \\ 
    \noalign{\smallskip}
  \end{tabular}
  \end{ruledtabular}
}\label{table:achieveda}

\subfloat[Maximum storage capacity and applied flux]{
  \begin{ruledtabular}
  \begin{tabular}{lcccc}
    \noalign{\smallskip}
    & \multicolumn{2}{c}{MAX COARSE $\Phi_0$} 
    & \multicolumn{2}{c}{Max Coupled Flux $(\Phi_0)$} \\     
    \noalign{\smallskip} \cline{2-5} \noalign{\smallskip}
    DAC Type    & \text{Designed} & \text{Achieved} & \text{Designed} & \text{Achieved} \\ 
    \noalign{\smallskip}\hline\noalign{\smallskip}
    Qubit Flux  &  17      &   22     &   0.050  &  0.077   \\ 
    CCJJ        &  17      &   22     &   0.093  &  0.086   \\ 
    L-Tuner     &  41      &   40     &   0.460  &  0.339   \\ 
    Coupler     &  41      &   35     &   0.960  &  0.665   \\
    \noalign{\smallskip}
  \end{tabular}
  \end{ruledtabular}
}\label{table:achievedb}
\end{table*}

\begin{figure}
\begin{center}
\includegraphics[width=3.3in]{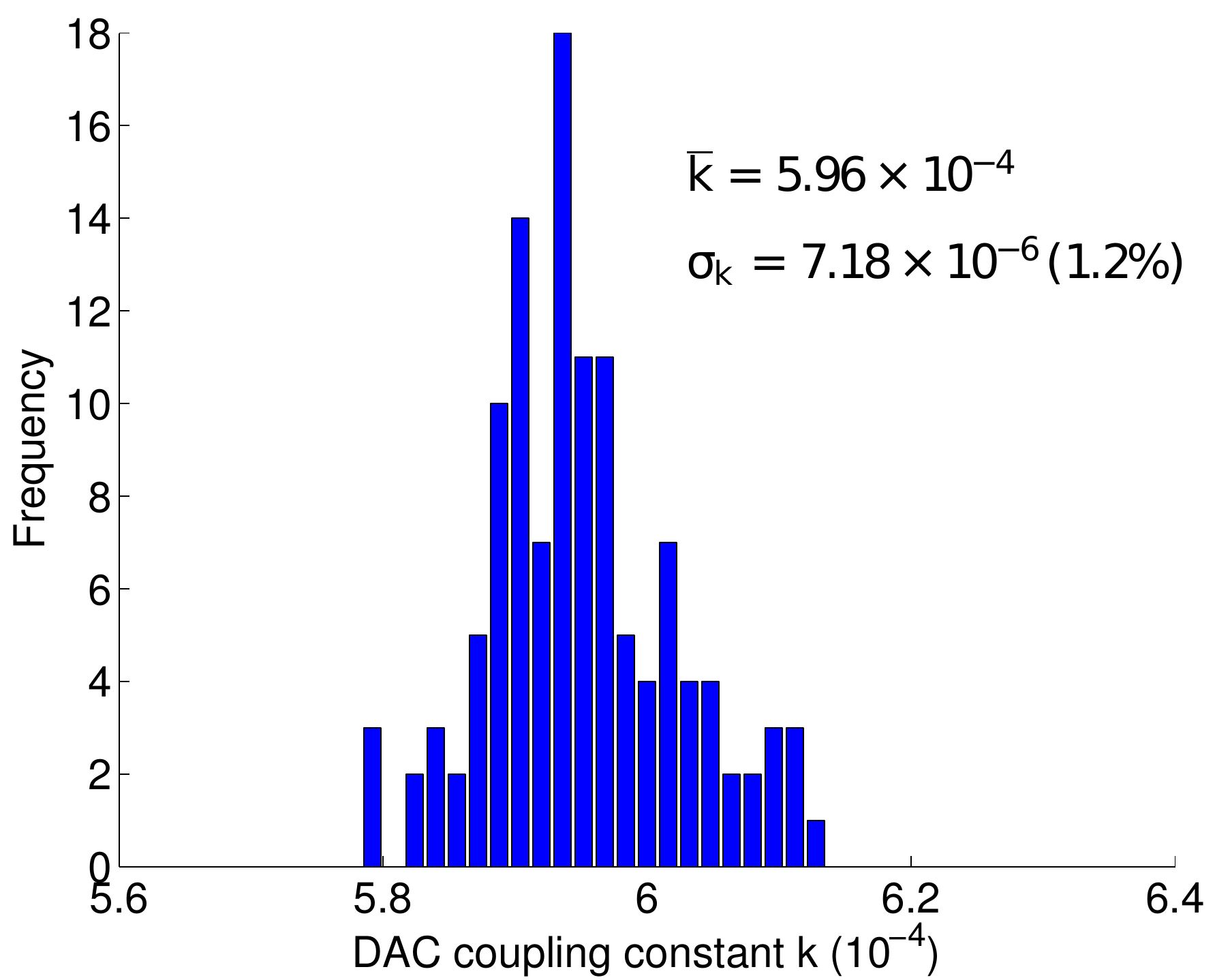}
\caption{Distribution of DAC $k$ values for 121 identically designed 
qubit flux bias DACs on one of the chips tested. 
The relative standard deviation is 1.2\%.}
\label{fig:dac-hist}
\end{center}
\end{figure}

Table~\ref{table:achieved}(a) summarizes design targets vs. achieved
COARSE and FINE step sizes for the various types of DAC implemented on
the eight-qubit block. Data from Table~\ref{table:achieved}(a) are
extracted from separate {\em break-out} versions of the
circuit. Uncertainties reported derive from the measurement
uncertainty of that parameter for that individual device.  Measuring
these parameters {\em in-situ} using the feedback technique described
above shows device-to-device variation with a standard deviation of
typically 2\%. For example, the distribution of $k$ values measured
for qubit flux bias DACs on a chip containing a 4x4 array of tiled
unit cells, shown in Fig.~\ref{fig:dac-hist}, exhibits a relative
standard deviation of 1.2\%. This is consistent with observed
variation in mutual inductance of simple microstrip transformers used
in this circuit, such as might occur with variations in dielectric
thickness of the same scale.  Table~\ref{table:achieved}(b) summarizes
the maximum number of SFQ and maximum coupled flux by each DAC type.

It is worth observing that we were able to confirm that \emph{all} 64 of
the two stage DACs on the single-unit-cell eight-qubit chips discussed
here yielded. By yielded we mean that they behaved as expected, per
Tables~\ref{table:achieved}(a) and \ref{table:achieved}(b), and that
variations in coupling between identically designed DACs were of the
order of 2\%.  Moreover, there were no significant differences in
maximum storage capacity between identically designed copies.  This
strongly suggests that the DAC storage coils yielded.  An inter- or
intra-layer short in one of the coils would likely have changed that
coil's storage capacity significantly, and this would have been
observable.

Deviations between design targets and achieved parameters for this
chip are as large as 30\% for some of the couplings.  This is
primarily due to the challenge of performing sufficiently accurate 3D
electromagnetic modeling of the superconducting inductors and
transformers used in creating the two-stage DACs.

\subsection{Demonstration of Demultiplexer Functionality}

Delivering pulses to the DAC requires that bias current and address
signals be applied to the demultiplexer tree. Bias current is shared
for all demultiplexer circuits in a particular tree, and address is
common for all demultiplexers at a particular level in the tree.  The
design and fabrication of the chip must be sufficiently uniform such
that all cells work with common levels. It is also necessary that the
operating margins are wide enough that a robust, low error rate
operating point can be obtained.

\begin{figure}[b]
\begin{center}
\includegraphics[width=3.3in]{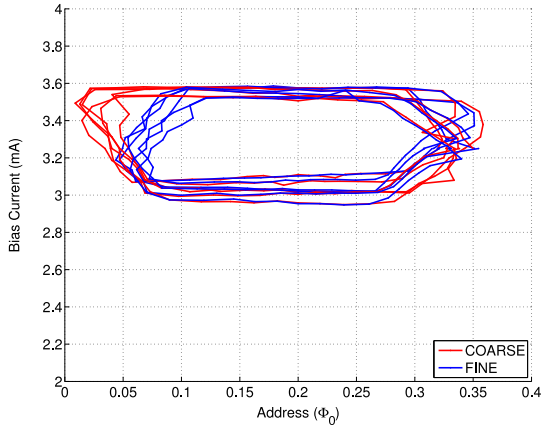}
\caption{Combined operating margins in global bias current current (for one
  address tree) vs. the magnitude of common address level for six
  coupler DACs from the same eight qubit circuit block. Two of these
  are addressed by one of the address trees on the chip, the rest by
  the other.}
\label{fig:margins}
\end{center}
\end{figure}

As discussed in section~\ref{sect:dac-dc-squid}, several DACs that are
attached to unused boundary couplers are wired up for use as dc-SQUIDs.
Operating margins required to address the FINE and COARSE DAC stages
of each of these were obtained with respect to global bias current and
level of address signal.  These operating regions are shown in
Fig.~\ref{fig:margins}.  Routing an SFQ pulse to a particular DAC
requires the successful navigation of six demultiplexer gates, each
with its own address current.  While the signs of these various
address levels may differ, their magnitude in flux was held to a
common value, and this common magnitude of address flux is the address
axis in Fig.~\ref{fig:margins}.

As far as addressing these six DACs, there is clearly adequate
uniformity in this demultiplexer tree that they can all be operated at
a common bias current and address level. We have determined that all 64 DACs
on the chip discussed here, as well as those on another subsequently
tested, were addressable with chip-wide common bias current and address
levels.

\begin{figure}
\begin{center}
\includegraphics[width=3.3in]{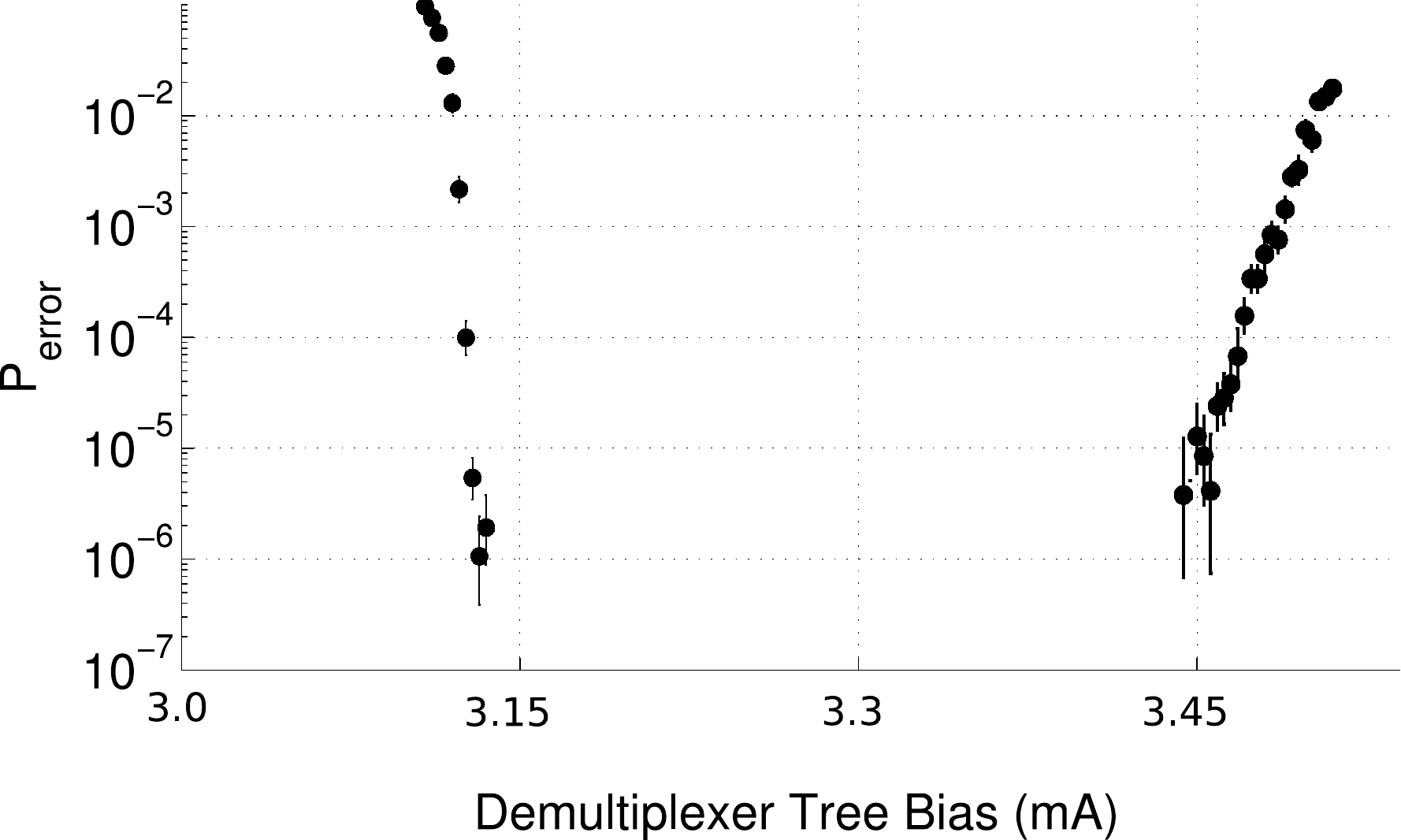}
\caption{Dependence of demultiplexer error rate on bias current current near
  nominal address levels. At the operating point, $P_{err}$ was bounded to
  be less than $2.5\times 10^{-7}$ with 95\% confidence.}
\label{fig:error}
\end{center}
\end{figure}

Fig.~\ref{fig:margins} shows the boundary outside of which the
probability of failing to increment a DAC stage is of order 0.1 or
higher. As mentioned in section~\ref{sect:require}, we require the
error rate to be considerably less than this.  The dependence of error
probability in SFQ circuits has been studied in some detail by a few
different groups \cite{herr96,herr99,satchell}.  However, we are
interested in the aggregate error probability of the entire
demultiplexer tree.  The probability that a pulse fails to be loaded
into a DAC was measured as a function of demultiplexer bias current at
nominal address level, and is shown in Fig.~\ref{fig:error}.

As expected, the margins decrease as the error probability requirement
decreases.  There is a significant bias current range with
$\mathrm{P}_{error}<10^{-6}$.  At the chosen operating point, over
15,000,000 operations were performed, and no errors were observed.
This places an upper bound on the probability of error $P_{err} <
2.5\times10^{-7}$ with 95\% confidence.

It is not sufficient that the address tree route pulses to the
addressed DAC. It must do so exclusively, and not route pulses to any
other DAC. Confirming that pulses arrived at the intended location, a
requirement to attain the data shown above, does not demonstrate
exclusivity. 

Indeed if significantly overbiased, with no address applied, the
demultiplexer tree is capable of operating in a {\em broadcast} mode,
where pulses are duplicated rather than routed at each 1:2
demultiplexer node. While this should not happen under normal
circumstances, failure to test for this would be an oversight.

Testing exclusivity was performed for most of the DACs on one of the
chips tested at nominal bias current and address levels by (1)
confirming that a DAC $\pmb{D}$ was addressable and could be
programmed, and then (2) sending pulses to {\em every other} DAC on
that tree and confirming afterward that the state of $\pmb{D}$ was
unchanged.  No cases of pulse misdirection were observed.

\section{Conclusion}
\label{sect:conc}

We have presented a description of a functioning system of on-chip
programmable magnetic memory designed to manipulate the parameters
and state of rf-SQUID superconducting qubits for use in implementing
AQO algorithms. The system is inherently scalable, and in turn allows
specialized AQO hardware to be scaled to very large numbers of
devices.  

Based on classical manipulation of single quanta of magnetic flux, the
system was implemented in a planarized four superconductor metal layer
process with $0.6\mu\text{m}$ minimum junction diameter and
$0.25\mu\text{m}$ lines and spaces for wiring. The control system was
fabricated on-chip, in the same process as the qubits and inter-qubit
couplers.

Both the two-stage flux DACs used to manipulate the various controls
on the qubits, and the demultiplexer address tree used to address
those DACs were shown to work as intended.  The address tree is shown
to pass SFQ pulses with very low error rate, and to address the DACs
exclusively.  The two-stage DAC design was shown to be effective at
providing a flux bias with a dynamic range in excess of eight bits of
precision (at dc).  The design targets on several variants of DAC were
presented, and while the variations between designed and achieved flux
coupled into target sometimes reached 30\%, this is close enough to
satisfy our current requirements.

While the control system described here was designed to operate an AQO
processor, it is probable that the devices described - programmable
flux DAC, programmable variable gain element, SFQ demultiplexer tree -
can be used to control other types of quantum information processors
implemented with superconductors.

\bibliography{0907.3757}

\end{document}